\documentclass{article}

\usepackage[OT1]{fontenc}

\usepackage{microtype}
\usepackage{graphicx}
\usepackage{subcaption}
\usepackage{booktabs} 
\usepackage{array}

\usepackage{xurl} 
\usepackage{hyperref}

\newcommand{\tautwo}{$\tau^2$-bench}

\usepackage[preprint]{icml2026}
\makeatletter
\renewcommand{\Notice@String}{}
\makeatother
\hypersetup{
  pdftitle={Repairs for Whom? Evaluator--Executor Mismatch in Selecting Repairs for Failed LLM Agents},
  pdfauthor={Yijun Liao and Fanwei Liang},
  pdfsubject={Research preprint}
}

\usepackage{amsmath}
\usepackage{amssymb}
\usepackage{mathtools}
\usepackage{amsthm}

\usepackage[capitalize,noabbrev]{cleveref}

\theoremstyle{plain}

\theoremstyle{definition}

\theoremstyle{remark}

\usepackage[disable,textsize=tiny]{todonotes}

\icmltitlerunning{Repairs for Whom? Evaluator--Executor Mismatch in Agent Repair}

\begin{document}

\twocolumn[
  \icmltitle{Repairs for Whom? Evaluator--Executor Mismatch\\in Selecting Repairs for Failed LLM Agents}

  \begin{icmlauthorlist}
    \icmlauthor{Yijun Liao}{independent}
    \icmlauthor{Fanwei Liang}{scut}
  \end{icmlauthorlist}

  \icmlaffiliation{independent}{Independent Researcher}
  \icmlaffiliation{scut}{South China University of Technology}

  \icmlkeywords{LLM agents, failure attribution, counterfactual replay, agent repair, evaluation}

  \vskip 0.3in
]



\printAffiliationsAndNotice{Code and data: \href{https://github.com/lyj20071013/repairs-for-whom}{github.com/lyj20071013/repairs-for-whom}.\newline}

\begin{abstract}
  When an LLM agent fails, repair methods choose which action to replace and what to put in its place. They often score candidate repairs with greedy replay, a stronger model, an environment oracle, or a judge that reads the failed log. We ask whether these evaluators pick repairs that help the agent that will actually execute them. Holding failed trajectories and candidate single-step repairs fixed, we label every candidate with independent rollouts of the deployed policy and measure how much deployment value each evaluator's choice loses. On failures not screened for repair value, pre-specified tests show that greedy replay loses 0.08--0.13 in expected net success gain in ALFWorld and 0.06--0.13 in net reward in WebShop, relative to selection with the deployed agent's own rollouts. At an equal rollout budget, selecting repairs for Qwen2.5-7B with Qwen2.5-14B rollouts loses 0.25 in ALFWorld and 0.07 in WebShop, twenty 14B rollouts per candidate are worth no more than two of the 7B agent's own, and fresh rollouts confirm that the two models prefer opposite repairs on selected disagreement cases. The agent's own fine-tuned successor is a worse evaluator still: after one round of training on verified repairs, its rollouts lose 0.55--0.60 when they select repairs for the agent it was trained from, as a pre-specified test confirms. Relative to what selection is worth, substitute models forgo 11--53\%; on $\tau^2$-bench, where selecting is worth only 0.10, they lose 0.03. Repairs should be verified with the executor that will run them, and repair labels should name their executor, including its training version.
\end{abstract}

\section{Introduction}
\label{sec:intro}

\begin{figure}[t]
  \centering
  \includegraphics[width=\columnwidth]{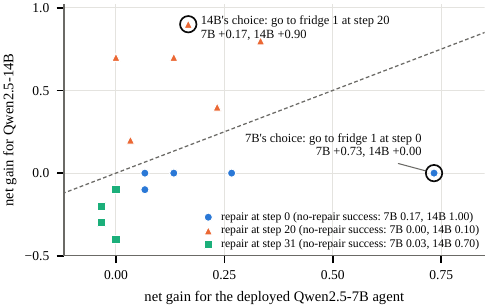}
  \caption{\textbf{Which repair is best depends on who executes it.} One failed ALFWorld trajectory of the deployed Qwen2.5-7B agent; each point is one of 18 candidate repairs at three steps, placed by its net gain (change in success probability over keeping the original action) when the 7B agent (30 rollouts per branch, horizontal) or Qwen2.5-14B (10 rollouts, vertical) executes the rest. The same action, going to the fridge, is the 7B agent's best repair at step~0 but the 14B model's at step~20; at step~0 the 14B model already succeeds without help, so the repair gains it nothing. Qwen3.8-flash succeeds from every branch (all net gains 0). An illustrative case from a screened pool.}
  \label{fig:example}
\end{figure}

LLM agents fail in long, multi-step interactions, and a growing line of work analyzes failed trajectories to find out what went wrong and how to fix it \citep{zhang2025whowhen,zhang2025agentracer,zhu2025agentdebug,ma2025dover,shah2026car,bonagiri2026causalflow,zhu2026aed}. Some methods identify the step at which an error was made or the agent responsible for it. Others choose a \emph{repair}---a step to change and an action to put there---or define their labels by which counterfactual correction would have turned the failure into a success \citep{zhang2025agentracer,ma2025dover,shah2026car,bonagiri2026causalflow}. This paper is about the second decision. Which repair is best is a counterfactual about the rest of the episode, and therefore depends on who executes that rest (\cref{fig:example}).

In practice, the executor imagined during evaluation is often not the agent that will run the repair. Deterministic replay is attractive for reproducibility; oracles and planners check whether the task is still solvable; verification relies on a handful of rollouts, for example accepting an intervention when two of three replays succeed \citep{ma2025dover}; corrections are proposed and validated by a stronger model with access to the ground truth \citep{zhang2025agentracer}; and LLM judges read the log and name the decisive step \citep{zhang2025whowhen}. Each of these substitutes an \emph{evaluator} for the deployed \emph{executor}, implicitly assuming that the repair that looks best to the evaluator is also best for the agent that will execute it.

That a repair's value depends on the executing policy is unsurprising in principle: value functions are policy-dependent, and step-level Monte-Carlo labels are known to depend on the policy that completes them \citep{xiong2024ipr,choudhury2025agentprm,meng2026pccc}. But a different value estimate need not change a decision: two evaluators can disagree on every candidate's value yet select the same repair. What matters for a repair method is \emph{decision regret}: the deployment value lost by the repair it actually selects. Whether mismatched evaluators incur such regret in practice, and how much, is not known.

We measure this regret directly. Our protocol holds fixed the failed trajectories of a deployed agent and a set of candidate single-step repairs, estimates every candidate's net gain for the deployed agent from rollouts that no selection uses, and compares each evaluator's choice with \emph{matched selection}, which uses the deployed agent's own rollouts. We apply it to a gridworld, ALFWorld \citep{shridhar2021alfworld}, WebShop \citep{yao2022webshop}, and \tautwo{} \citep{barres2025tau2}, with three deployed agents whose failures we repair (Qwen2.5-7B-Instruct, Llama-3.1-8B-Instruct, and Qwen3.5-2B) and failure pools that were not screened for repair value, under tests pre-specified in a dated analysis plan; stronger models serve only as substitute executors.

Substitutes cost the deployed agent real value. On the unscreened pools, greedy replay of the deployed agent loses 0.08--0.13 in success probability in ALFWorld and 0.06--0.13 in reward in WebShop, and choosing with five rollouts per candidate from Qwen2.5-14B instead of the deployed 7B agent loses 0.25 in ALFWorld and 0.07 in WebShop; between Qwen2.5-7B and Llama-3.1-8B the loss is present in both directions in both environments. These losses are best read against what selection is worth. On \tautwo{}, where matched selection beats random selection by only 0.10, a 4B model's choices cost the 2B agent 0.03; across pools, substitute models forgo 11--53\% of the value of selecting (\cref{sec:main}).

More rollouts of a substitute close the gap slowly: twenty 14B rollouts per candidate select repairs worth about as much to the 7B agent as one of its own in ALFWorld. Fresh rollouts confirm that the 7B and 14B models prefer opposite repairs on selected disagreement cases, whereas Llama mainly fails to separate the repairs the 7B agent needs (\cref{sec:why}). The agent's own successor fails in the same way, and more severely. After one round of fine-tuning on verified repairs, a successor's rollouts lose 0.55--0.60 when they select repairs for the agent it was trained from, because they show little preference between repairs that its predecessor tells apart sharply (\cref{sec:training}).

The loss lies in choosing among a trajectory's positions and candidates. The executor that scores the counterfactuals changes where a repair-location label points and which selection method a benchmark ranks first, yet where one candidate is good for every executor, per-position labels chosen by another model are nearly as good as the agent's own (\cref{sec:consequences}). A few rollouts of the deployed agent per candidate recover most of the loss (\cref{sec:method}). Our contributions are a measurement protocol for repair selection under evaluator--executor mismatch; evidence from three environments and a bound from a fourth, including preference reversals on fresh rollouts and a mismatch between an agent and its own fine-tuned successor; and the consequences for repair labels, benchmarks, and low-cost verification. We release code, trajectories, all rollout outcomes, and the analysis plan.

\section{Problem Setup}
\label{sec:problem}

An agent acts in an environment with a step budget $T$. A trajectory $\tau=(o_0,a_0,o_1,a_1,\dots)$ alternates observations and actions, and $h_t$ denotes the history before action $a_t$. We study trajectories that failed. A \emph{single-step repair} $r=(t,a')$ replaces $a_t$ with an alternative action $a'$ at history $h_t$; the rest of the episode is then produced by an \emph{executor} $e$. An executor is more than a model name: it comprises the model weights, the prompt and history protocol, the decoding parameters, and the rules for handling invalid actions. Let $p_e(h_t,a)$ be the probability that the episode succeeds within the remaining budget when $a$ is taken at $h_t$ and $e$ executes the rest (in WebShop, the expected reward in $[0,1]$). The \emph{net gain} of a repair under executor $e$ is
\begin{equation}
  R_e(r) \;=\; p_e(h_t,a') \;-\; p_e(h_t,a_t).
  \label{eq:netgain}
\end{equation}
We use the net gain rather than the post-repair success probability $p_e(h_t,a')$ as the primary objective because the original action can itself succeed on a rerun, and a repair should be credited only for the success it adds. Across positions the baselines $p_e(h_t,a_t)$ differ, so the two objectives can rank repairs differently; we report the post-repair objective in \cref{app:post}.

The repair will be executed by the deployed executor $e_0$, which in our experiments is the policy that produced the failure. Given the candidate set $\mathcal{R}_\tau$ of a trajectory, an \emph{evaluator} $c$ assigns each candidate a score and selects $\hat r_c\in\arg\max_{r\in\mathcal{R}_\tau} S_c(r)$; ties are broken uniformly at random, and we average over them. The \emph{deployment value} of this choice is $R_{e_0}(\hat r_c)$. An evaluator can depart from the deployed executor in two distinct ways (\cref{tab:evaluators}). It can change the \emph{target}, estimating a different quantity than $R_{e_0}$: the success of greedy decoding, the solvability of the task, the success of another model, or a judge's opinion. Or it can keep the target but cut the \emph{budget}, estimating $R_{e_0}$ from a handful of rollouts. We report the two separately, so that ``more samples help'' is not mistaken for evidence of mismatch.

As the reference we use \emph{matched selection} $\hat r_\star$, which estimates $R_{e_0}$ for every candidate from the deployed executor's own rollouts. The \emph{selection loss} of evaluator $c$ is
\begin{equation}
  L_c \;=\; \mathbb{E}_\tau\!\left[\,R_{e_0}(\hat r_\star)-R_{e_0}(\hat r_c)\,\right],
  \label{eq:loss}
\end{equation}
where both deployment values are estimated from rollouts that neither selection has seen. Matched selection is itself a finite-sample procedure, not the true best repair, so $L_c$ compares two selection procedures and can be negative. Because both are scored on the same independent rollouts, the comparison needs no validation-set maximum and is free of the winner's curse. As a positive control, matched selection must beat random selection by more than our margin $\delta=0.10$; otherwise the candidates offer too little selection gain for a small loss to be informative, and we would mark the pool as uninformative.

\begin{table}[t]
  \caption{Evaluators compared against matched selection. \emph{Target}: the evaluator estimates a different quantity than the deployed executor's net gain. \emph{Budget}: same target, fewer samples or a decision rule applied to the deployed executor's own rollouts. The published decision rules re-implement only the verification step of each method within our fixed-candidate protocol; they are not full reproductions.}
  \label{tab:evaluators}
  \centering
  \footnotesize
  \setlength{\tabcolsep}{3pt}
  \begin{tabular}{@{}p{0.45\columnwidth}p{0.335\columnwidth}l@{}}
    \toprule
    Evaluator & Practice it stands for & Departs in \\
    \midrule
    Deployed executor, 1--5 rollouts & few-rollout verification & budget \\
    Same model at temperature 0.3 & close decoding proxy & target \\
    Greedy replay (temperature 0) & deterministic replay & target \\
    Solvability oracle $F$ & oracle / planner checks & target \\
    Oracle, ties broken by slack & strongest environment signal & target \\
    Planner action at earliest position & planner-based repair & target \\
    2-of-3 acceptance \citep{ma2025dover} & intervention verification & budget \\
    Single-replay flip \citep{bonagiri2026causalflow} & counterfactual repair & budget \\
    Earliest single success \citep{zhang2025agentracer} & earliest sufficient step & budget \\
    Latest significant step \citep{shah2026car} & point of commitment & budget \\
    Other models' rollouts (14B, Llama, flash) & stronger or different model & target \\
    LLM judge (log only / informed) & judge-based attribution & target \\
    \bottomrule
  \end{tabular}
\end{table}

\section{Experimental Setup}
\label{sec:exp}

We use four text environments in which an episode can be restored exactly to any intermediate history by replaying its prefix. The \emph{gridworld} is a fully observed $5\times5$ world with a key, a locked door, a gem, and a trap that ends the episode. We compute exactly whether the task is still solvable from any history ($F(h)\in\{0,1\}$) and how many steps of slack remain; stepping on the trap or wasting too many steps creates an \emph{irreversible point}. \emph{ALFWorld} \citep{shridhar2021alfworld} is a partially observed household environment with a 50-step budget and no irreversible actions; a Fast Downward planner \citep{helmert2006fastdownward} with access to the full state gives solvability and slack. \emph{WebShop} \citep{yao2022webshop} is a simulated online store in which the purchase ends the episode with a reward in $[0,1]$; its primary objective, fixed before any formal run, is reward. In \emph{\tautwo{}} \citep{barres2025tau2} the agent serves a simulated user in a retail or airline domain by calling tools on a database, and we score every rollout against the original task. \Cref{app:setup} gives the configurations.

The deployed agents, whose failures we repair and whose success defines deployment value, are Qwen2.5-7B-Instruct \citep{qwen2025qwen25}, Llama-3.1-8B-Instruct \citep{grattafiori2024llama3}, and, on \tautwo{}, Qwen3.5-2B \citep{qwen2026qwen35}, all served with vLLM \citep{kwon2023vllm} at temperature 0.7. Other executors---Qwen2.5-14B-Instruct, Qwen3.8-flash \citep{qwen2026flash}, Qwen3.5-4B on \tautwo{}, and each deployed model on the other's pool---use exactly the same prompt, history protocol, decoding parameters, and action parsing as the deployed agent. The LLM judge is Qwen3.8-flash with reasoning: it reads the trajectory and the candidate list, is told which model will execute the repair, and names one candidate.

\begin{table}[t]
  \caption{\emph{Unscreened}: the first failures of fresh episodes, not screened by repair value. \emph{Screened}: failures in which screening rollouts showed some alternative helping. Candidates (mean per position): L, proposals from LLMs (two from each of three; on \tautwo{}, up to three from Qwen3.5-4B, keeping only actions a deployed agent could produce); P, privileged action (the planner in ALFWorld); R, random action; C, every clickable element. Rollouts per stochastic branch for each executor, plus one greedy rollout each; in the gridworld, also 20 rollouts of the 7B agent at temperature 0.3.}
  \label{tab:pools}
  \centering
  \scriptsize
  \setlength{\tabcolsep}{3pt}
  \begin{tabular}{@{}llll@{}}
    \toprule
    Pool (deployed) & Failures & Candidates & Rollouts per branch \\
    \midrule
    \multicolumn{4}{@{}l}{\emph{Unscreened (primary)}} \\
    ALFWorld, 7B & 60/82 & L, P, R (6.1) & 7B 10, 14B 20, Llama 10 \\
    ALFWorld, Llama & 60/109 & L, P, R (6.0) & Llama 10, 7B 10 \\
    WebShop, 7B & 60/72 & C, L, P, R (11.5) & 7B 10, 14B 20, Llama 10 \\
    WebShop, Llama & 60/86 & C, L, P, R (10.1) & Llama 10, 7B 10 \\
    \tautwo{}, 2B & 115 & L, P (2.6) & 2B 10, 4B 6 \\
    \midrule
    \multicolumn{4}{@{}l}{\emph{Screened (controlled comparisons)}} \\
    Gridworld, 7B & 34/107 & all legal (6) & 7B 40, 14B 10, flash 10 \\
    ALFWorld, 7B & 30/78 & P, R (6) & 7B 30, 14B 10, flash 10 \\
    \bottomrule
  \end{tabular}
\end{table}

\Cref{tab:pools} lists all pools. The \emph{unscreened} pools take failures of fresh episodes without looking at any repair signal: the first 60 in collection order in ALFWorld and WebShop, and on \tautwo{} 115 failed conversations of the 2B agent across 96 tasks, fixed by a hash. Repairs are placed only at positions that leave room to recover. The \emph{screened} pools, collected first, keep only failures in which screening rollouts, never reused, showed some single-step repair helping; we use them as a controlled setting in which to compare many evaluators and to analyze mechanisms.

Positions are fixed before candidates are generated: the irreversible point and three earlier positions in the gridworld, one position from each third of the eligible steps in ALFWorld and WebShop, and positions before the first erroneous write on \tautwo{}. In the unscreened pools, candidates combine alternatives proposed by LLMs that read the failed log, a privileged action computed from hidden state (the planner's next action in ALFWorld, the next action toward the target product in WebShop, the reference next action on \tautwo{}), and a random action; WebShop adds every clickable alternative. Privileged and exhaustive candidates keep the candidate sets from reflecting any single proposer's preferences.

The unit of analysis is the trajectory. Matched selection is cross-fitted: candidates are selected on one fold of the deployed agent's rollouts and scored on the other (in the unscreened pools, five of ten rollouts select and five evaluate, averaged over 50 random splits); any evaluator that uses the deployed agent's rollouts follows the same split, and all other evaluators are scored on all of the deployed agent's evaluation rollouts. Comparisons with other executors use the same number of rollouts per candidate on both sides. We report 90\% intervals from 2,000 bootstrap replicates that resample trajectories, and one-sided tests at $\alpha=0.05$, Holm-corrected within each pre-specified family. Intervals are percentile intervals on the unscreened pools (resampling tasks on \tautwo{}) and bias-corrected two-level intervals on the screened pools (\cref{app:stats}). Designs, primary tests, decision thresholds, and predictions were pre-specified in a dated analysis plan before the corresponding data were collected; analyses added later are labeled exploratory, and \cref{app:audit} gives the status of every analysis.

\section{Substitute Evaluators Choose Worse Repairs}
\label{sec:main}

\begin{figure}[t]
  \centering
  \includegraphics[width=\columnwidth]{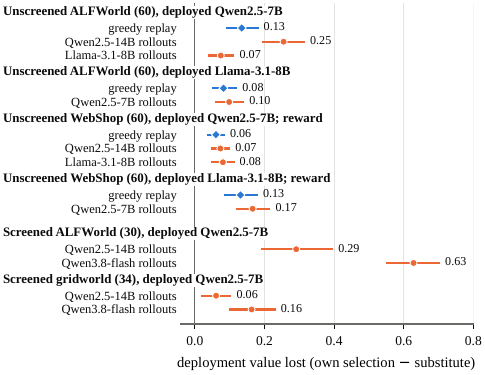}
  \caption{\textbf{Deployment value lost by substitute evaluators.} Each row is a paired difference: the deployed agent's value of repairs selected with its own rollouts minus that of the substitute's, both scored on its held-out rollouts. Unscreened pools (every pre-specified primary test in ALFWorld and WebShop, all Holm-confirmed): own selection with five rollouts per candidate versus greedy replay (diamonds) or five rollouts of another model (circles); nonparametric 90\% intervals. Screened pools: ten rollouts per candidate on both sides; bias-corrected 90\% intervals. ALFWorld and gridworld values are success probabilities, WebShop values rewards.}
  \label{fig:forest}
\end{figure}

\paragraph{Unscreened failures.}
\Cref{fig:forest} shows every primary test on the unscreened ALFWorld and WebShop pools. The positive control passes in each, and all ten pre-specified primary tests are confirmed after Holm correction (\cref{tab:registered}). Greedy replay of the deployed agent loses 0.08--0.13 in ALFWorld and 0.06--0.13 in WebShop. The largest loss comes from a stronger model: selecting with Qwen2.5-14B's rollouts instead of the 7B agent's own, five per candidate on both sides, loses 0.25 [0.19, 0.32] in ALFWorld, but 0.07 [0.05, 0.10] in WebShop, no more than Llama's selections lose there. Between the similar-sized Qwen and Llama models the loss is present in both directions in both environments: 0.07--0.10 in ALFWorld, and in WebShop 0.08 when Llama selects for the 7B agent and 0.17 when the 7B model selects for Llama.

The losses do not come from the privileged candidates: restricted to candidates proposed by at least one LLM (exploratory), greedy replay loses 0.11 [0.07, 0.16] and 14B selection 0.20 [0.14, 0.27] in ALFWorld, and the WebShop losses are 0.04--0.13. The qualitative conclusions persist under the post-repair success objective in ALFWorld (greedy 0.10, 14B 0.29) and under WebShop's secondary success readout, which is weaker (0.03--0.09; \cref{app:post,app:webshopres}). On 100 unscreened gridworld failures the losses are smaller (greedy replay 0.05, solvability oracle 0.07), because many such failures admit no useful single-step repair (\cref{app:gridpool}).

\paragraph{Screened failures: many evaluators.}
The screened pools allow a wider range of evaluators (\cref{tab:screened} in \cref{app:screened}). With the target held fixed, a few rollouts suffice: three per candidate lose 0.11 in the gridworld and 0.07 in ALFWorld, ten are within noise of matched selection, and the same model at temperature 0.3 stays within our 0.10 margin, so not every deviation from the deployed executor hurts. The substantial losses come from evaluators that estimate something other than the deployed agent's success. At equal budgets of ten rollouts per candidate, other executors lose 0.29 [0.19, 0.40] (14B) and 0.63 [0.55, 0.70] (Qwen3.8-flash) in ALFWorld and 0.06 [0.02, 0.11] and 0.16 [0.10, 0.23] in the gridworld (\cref{fig:forest}); on these pools the most capable executor lost the most.

\paragraph{What the substitutes get wrong.}
Greedy replay loses 0.23 in the gridworld and 0.18 in ALFWorld, in part because greedy decoding repeats its own mistakes: switching to sampling whenever the greedy agent revisits a state recovers 0.10 of the gridworld loss. Environment signals know \emph{what} to change but not \emph{where}. In 29 of 30 screened ALFWorld trajectories the best candidate is the planner's action, yet the planner's action at the earliest position loses 0.14 and solvability with slack tie-breaking 0.28, because where the deployed agent benefits most depends on how it behaves afterwards; across a trajectory's three positions, the best net gain differs by 0.51 on average. Published acceptance rules applied to the deployed agent's own rollouts---two of three replays succeed \citep{ma2025dover}, a single replay flips the outcome \citep{bonagiri2026causalflow}, or the earliest single success \citep{zhang2025agentracer}---do not rank what they accept and lose 0.23--0.58. A reasoning judge that reads the log loses 0.06 in the gridworld and 0.53 in ALFWorld, where the log does not reveal what closed containers hold; told their contents, it finds the planner's action but the best position in only 47\% of trajectories, and still loses 0.19. \Cref{app:screened} gives the details and a human reader's choices.

\begin{table}[t]
  \caption{Equal-budget selection loss against what selection is worth (matched minus random selection), five rollouts per candidate on both sides. Rows above the rule use the pre-specified statistics of each pool; the shares are post hoc, with 90\% intervals that resample trajectories. Successor rows give the range over four successors (\cref{sec:training}).}
  \label{tab:ratio}
  \centering
  \footnotesize
  \setlength{\tabcolsep}{3pt}
  \resizebox{\columnwidth}{!}{%
  \begin{tabular}{@{}llccc@{}}
    \toprule
    Pool & Deployed $\leftarrow$ selector & Loss & Worth & Share \\
    \midrule
    ALFWorld & 7B $\leftarrow$ 14B & 0.255 & 0.707 & 0.36 [0.28, 0.44] \\
    ALFWorld & 7B $\leftarrow$ Llama & 0.074 & 0.707 & 0.11 [0.05, 0.16] \\
    ALFWorld & Llama $\leftarrow$ 7B & 0.098 & 0.737 & 0.13 [0.08, 0.19] \\
    WebShop & 7B $\leftarrow$ 14B & 0.073 & 0.163 & 0.45 [0.33, 0.56] \\
    WebShop & 7B $\leftarrow$ Llama & 0.081 & 0.163 & 0.49 [0.32, 0.66] \\
    WebShop & Llama $\leftarrow$ 7B & 0.166 & 0.314 & 0.53 [0.41, 0.65] \\
    \tautwo{} & 2B $\leftarrow$ 4B & 0.026 & 0.099 & 0.26 [0.04, 0.46] \\
    \midrule
    ALFWorld & 7B $\leftarrow$ its successors & 0.55--0.60 & 0.711 & 0.77--0.84 \\
    ALFWorld & successors $\leftarrow$ 7B & 0.02--0.05 & 0.10--0.17 & --- \\
    \bottomrule
  \end{tabular}}
\end{table}

\paragraph{\tautwo{}: a pool with little to select.}
On \tautwo{}, with Qwen3.5-2B deployed and Qwen3.5-4B as the substitute, matched selection beats random selection by only 0.10 [0.07, 0.13]. That is below our positive control, so by our own rule this pool can neither confirm nor rule out mismatch. It does bound it: selecting with five 4B rollouts instead of the 2B agent's own loses 0.03 [0.01, 0.05], and greedy replay loses 0.02 [0.00, 0.03]. Selection is worth little here because there is little to repair. From these positions the original action already succeeds in half of the 2B agent's rollouts, and even the best candidate in hindsight adds only 0.28 (\cref{app:tau2}).

\paragraph{Loss relative to the value of selection.}
\Cref{tab:ratio} divides each equal-budget loss by the value of selecting at all. Absolute losses differ tenfold across pools. As a share of the selection value, substitute models forgo 11--53\%. WebShop, whose absolute losses are small, has the largest shares, and \tautwo{} falls inside the range of the other pools, although its interval is wide. The share is not a constant of mismatch: for the 7B agent in ALFWorld it is 0.36 when the 14B model selects and 0.11 when Llama does. A small absolute loss may therefore only mean that the pool leaves little to choose.

\section{More Samples, or Different Preferences?}
\label{sec:why}

A substitute executor's choices could be worse merely because they are noisy estimates of the deployed agent's values, in which case more of its rollouts would close the gap; or because the two executors benefit from different repairs.

\begin{figure}[t]
  \centering
  \includegraphics[width=\columnwidth]{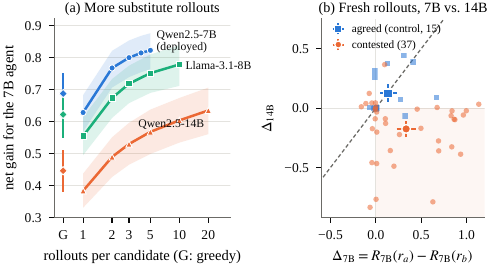}
  \caption{\textbf{More rollouts do not fix the executor, and preferences reverse on fresh rollouts.} (a) Unscreened 7B ALFWorld pool: net gain for the 7B agent of repairs selected with $k$ rollouts per candidate from each executor (the deployed agent's own selection is cross-fitted, hence $k\le5$); bands and bars: 90\% intervals. (b) Fresh rollouts (30 per branch and executor) on the 37 trajectories in which the 7B and 14B models' best candidates differed in earlier rollouts ($r_a$: 7B's, $r_b$: 14B's) and on 15 control trajectories in which they agreed ($r_b$: 7B's second best); small marks are trajectories, large marks group means with 90\% intervals. Shaded: opposite preferences; dashed: equal preference.}
  \label{fig:curves}
\end{figure}

\paragraph{More substitute rollouts help slowly.}
\Cref{fig:curves}a shows, on the unscreened 7B ALFWorld pool, the deployment value of repairs selected with $k$ rollouts per candidate. The 14B model's selections improve from 0.38 with one rollout to 0.63 [0.56, 0.71] with twenty, about the value of selecting with a single deployed-agent rollout per candidate (0.63), and below the deployed agent's greedy replay (0.69) or five of its rollouts (0.82). On the same GPU, twenty 14B rollouts per candidate cost 31--39 times the time of that one-rollout selection. Llama's selections come closer but need ten rollouts per candidate to match two of the 7B agent's own (0.78 versus 0.77). In WebShop, twenty 14B rollouts or ten Llama rollouts per candidate are worth between one and two of the 7B agent's own, at the GPU time of about 21 of them (\cref{app:budget}). More rollouts of the wrong executor are a poor substitute for a few of the right one.

\paragraph{Saturation does not explain all of the loss.}
A capable executor may already succeed where the deployed agent needs help, leaving its net gains near zero exactly where they matter. In screened ALFWorld, Qwen3.8-flash succeeds without any repair at the position of the 7B agent's best repair in 77\% of trajectories, so its rankings carry little information there (a post-hoc description). For the 14B model we pre-specified a stratified check on the unscreened pool: in the 41 of 60 trajectories in which the 14B model, keeping the original action at the position of the 7B agent's best repair, succeeds less than half the time, the equal-budget loss is still 0.17 [0.11, 0.24] (0.54 in the 10 trajectories where it succeeds at least 90\% of the time). Saturation increases the loss but is not sufficient to explain it. Llama falls in the unsaturated stratum in 55 of 60 trajectories, with a loss of 0.06 [0.03, 0.10].

\paragraph{Fresh rollouts confirm opposite preferences.}
Because a disagreement in the original rollouts could itself be selection noise, we pre-specified a check that uses only fresh rollouts on every unscreened 7B trajectory in which the 7B and 14B models' best candidates differed (37 trajectories; \cref{fig:curves}b). With $r_a$ the 7B agent's earlier best candidate, $r_b$ the 14B model's, and $\Delta_e=R_e(r_a)-R_e(r_b)$ the fresh preference of executor $e$ for $r_a$, the 7B agent still prefers its own choice ($\Delta_{\mathrm{7B}}=0.33$ [0.23, 0.44]) and the 14B model still prefers its own ($\Delta_{\mathrm{14B}}=-0.18$ [$-$0.25, $-$0.11]). The two preferences are individually significant and opposite in 8 trajectories and never significantly aligned. In the 15 control trajectories, $\Delta_{\mathrm{7B}}-\Delta_{\mathrm{14B}}$ is 0.01 [$-$0.08, 0.11], against 0.51 [0.38, 0.64] in the contested ones. Because the contested trajectories were selected with earlier rollouts, this shows that genuine reversals exist, not how common they are.

The analogous check for the 7B--Llama pair (34 contested 7B trajectories, run because its pre-specified trigger was met) replicates the 7B agent's preference (0.18 [0.10, 0.27]) but not Llama's ($-$0.02 [$-$0.07, 0.03]). Llama shows no stable opposing preference on average (\cref{tab:reversal}): its loss comes mainly from failing to separate the repairs the 7B agent needs. Two pre-specified cross-pool comparisons agree: Llama's selections cost the 7B agent on both ALFWorld pools (0.07 on the 7B pool, 0.10 on Llama's), whereas the 7B model's selections cost Llama only on Llama's own failures (0.10, against 0.02 [$-$0.00, 0.04]).

\section{The Same Model, After Training}
\label{sec:training}

Self-improvement pipelines verify repairs with the current policy, train on them, and carry on with the trained policy \citep{song2024eto,yuan2025agentr}. Is a fine-tuned model still the same executor when repairs are chosen?

We fine-tuned the 7B agent for one round on repairs that it had itself verified: LoRA \citep{hu2022lora} on 558 repair labels from training episodes disjoint from the pool, mixed with as many of the agent's own steps from successful episodes, under two seeds. Because most of those labels are the planner's action, we then trained a second pair on labels chosen without the planner's candidates, after pre-specifying its test (\cref{app:training}). All four ran ten rollouts on every branch of the unscreened 7B ALFWorld pool. Continuing on their own from the pool's positions, the first pair succeeds 0.46 and 0.36 of the time and the second 0.31 and 0.21, against 0.12 for the original agent. The first pair's analyses are post hoc.

\paragraph{The successor is a poor evaluator for its predecessor.}
Selecting repairs for the original 7B agent with five rollouts per candidate from a successor loses 0.60 [0.53, 0.67] and 0.59 [0.52, 0.65] for the first pair, and 0.56 [0.48, 0.63] and 0.55 [0.48, 0.62] for the second, confirming its pre-specified test for both seeds. This is 77--84\% of what selection is worth and more than twice the 14B model's loss on the same pool (\cref{tab:ratio}). The loss is not confined to positions a successor has outgrown: in the trajectories in which the successor, keeping the original action at the position of the 7B agent's best repair, succeeds less than half the time, the loss is still 0.48--0.55. Where the choices differ, held-out rollouts show the 7B agent preferring its own choice by 0.61--0.66, whereas the successors show little average preference between the two ($-0.06$ to $-0.01$; three of the four intervals include zero). A successor's average preference is far weaker than its predecessor's: like Llama, it mostly fails to separate the repairs the 7B agent needs, at a much higher cost.

\paragraph{The predecessor's labels are mostly no longer needed.}
In the other direction little is left to select: for the successors, matched selection beats random selection by 0.10--0.17, clearing our positive control for one of the four, and the original agent's choices cost them 0.02--0.05, about what the other training seed's choices cost. Position by position, 73--82\% of the positions at which the 7B agent needed a repair no longer need one for a successor, far more than rollout noise produces. Across the same positions, the successor's own best candidate gains nothing over the old label ($-0.02$ to 0.00). Old labels expire because positions stop needing repair, not because better actions appear (\cref{app:training}).

\paragraph{Version is part of the executor.}
One round of fine-tuning makes a model a different executor for the purpose of choosing repairs, and the direction matters: the stronger version of an agent is a worse judge of what the current version needs than a larger model (0.25) or a model from another family (0.07). Labels verified before an update should be checked for whether their positions still need repair before they are reused, which costs one branch per position. Such checks need a same-policy baseline: on \tautwo{}, a successor trained in the same way did not measurably change, yet 48\% of its predecessor's labels failed to replicate in its rollouts, no more than the 50\% that fail between two halves of the predecessor's own (\cref{app:training}).

\section{Consequences for Repair Labels and Benchmarks}
\label{sec:consequences}

Several failure-attribution benchmarks define their labels by counterfactual replay with a particular model---the earliest step whose correction suffices for success, or the step at which an injected error becomes decisive \citep{zhang2025agentracer,liu2026whowhenpro,zhu2026aed}. Such labels measure a repair's value for the replaying executor. We ask how much they, and method rankings built on them, depend on that executor (exploratory; screened pools).

For each trajectory, let the \emph{repair-location label} be the position of the candidate with the highest net gain---where an intervention helps most, not necessarily where the error was made. In ALFWorld, the position chosen by the 14B model's rollouts agrees with the deployed agent's in 32\% [26, 38] of trajectories and Qwen3.8-flash's in 29\% [25, 34], against 62\% [51, 72] between two disjoint halves of the deployed agent's own rollouts. In the gridworld, where the trap fixes the position, the labels agree as often as the deployed agent agrees with itself (72\% versus 66\%).

We scored 17 selection methods with each executor's rollouts (disjoint from those used for selection), as a benchmark would if it defined ground truth by that executor's counterfactuals. In ALFWorld, the rankings under 14B and Qwen3.8-flash scoring correlate only moderately with the ranking under the deployed agent (Kendall $\tau=0.51$ and 0.46; gridworld 0.77 and 0.60), and each executor ranks first the method that uses its own rollouts, which ranks 12th and 14th under 7B scoring. Such labels and rankings are legitimate measures of value \emph{for an executor}, but they should state which executor they are relative to; releasing every candidate's rollout outcomes for each executor, as we do, makes the dependence checkable.

\paragraph{What does transfer.}
Mismatch costs most when a method must choose \emph{among} positions and candidates. Suppose instead that every position is labeled with its best candidate whenever that candidate gains at least 0.20 (post hoc; \cref{app:labels}). In ALFWorld, the labels chosen by the 14B model or by Llama coincide with the 7B agent's own in 86--90\% of cases and are worth 0.68 and 0.73 to it, against 0.72 for its own: one privileged action is the best candidate at most positions, and every executor finds it. In WebShop, where no candidate dominates, another model's labels are worth 0.10 in reward against 0.17 for the agent's own, and 0.23 against 0.32. The fine-tuned successors label fewer than half of the 150 positions at which the 7B agent needs a repair, and their labels are worth 0.29--0.35 to it. A label for a given position transfers when the candidate set contains an action that is good for every executor; where to intervene does not transfer.

\section{When Is Deployed-Agent Verification Cheap?}
\label{sec:method}

The direct remedy is to verify candidates with a few rollouts of the deployed agent. We ask when this is cheap, counting the deployed-agent rollouts used for selection.

A few rollouts suffice when good candidates are present.
Three deployed-agent rollouts per candidate lose 0.11 and 0.07 on the screened pools (relative to 20 and 15 rollouts) and 0.02 on the unscreened 7B ALFWorld pool (relative to five), whereas greedy replay, at a third of that budget, loses 0.23, 0.18, and 0.13; in WebShop, three rollouts reach most of what five reach (\cref{app:method}).

A privileged signal helps when it knows what to change.
Environment signals tend to identify \emph{what} to change, and the deployed agent's rollouts reveal \emph{where}. A screen exploits this: at each position, keep only the candidate preferred by a privileged signal, run $k$ deployed-agent rollouts on each survivor and on the original action, and select the survivor with the highest estimated net gain. On screened ALFWorld, ten rollouts per survivor come within 0.009 of matched selection with 19\% of its rollouts. On the unscreened pools, five rollouts per survivor lose 0.04 [0.00, 0.07] on the 7B pool (pre-specified prediction: below 0.05) and 0.04 on the Llama pool, whereas keeping Qwen3.8-flash's first proposal instead loses 0.55 and 0.37. The screen works there because the planner's action is the best candidate in 86\% of unscreened 7B trajectories. In WebShop the privileged action is the best candidate in only 29\% and 20\% of trajectories, and the screen is no better than an LLM's first proposal on the 7B pool and worse than greedy replay on the Llama pool (\cref{app:method}).

Without a privileged signal, candidates are the bottleneck.
When an LLM both names the step and proposes two alternatives there, verifying them with six deployed-agent rollouts per trajectory gains 0.32 [0.22, 0.40], against 0.22 for applying the first proposal unverified and 0.64 for matched selection over LLM-proposed candidates at three fixed positions. The gap lies in the candidates: the better of the two proposals in hindsight is worth only 0.37 (exploratory; \cref{app:method}).

\section{Related Work}
\label{sec:related}

\paragraph{Failure attribution and repair.}
Who\&When \citep{zhang2025whowhen} introduced identifying the agent and step responsible for a multi-agent failure; Who\&When Pro \citep{liu2026whowhenpro} injects errors into successful runs, so the decisive step is defined by replay with the same system. AgenTracer \citep{zhang2025agentracer} labels the earliest step whose correction, proposed by a model that sees the ground truth, suffices for success. AgentDebug \citep{zhu2025agentdebug} and the Agent Error Dataset \citep{zhu2026aed} annotate root-cause errors at scale and re-execute from the critical step. AgentRx \citep{barke2026agentrx}, TrajDebug \citep{qi2026trajdebug}, and other methods \citep{xia2026agentlocate,nguyen2026soap} localize critical steps without executing a correction. These works evaluate localization or a single correction; we label a fixed set of candidate repairs with the deployed policy's success and ask which evaluator selects the right one.

\paragraph{Counterfactual replay and verification.}
DoVer \citep{ma2025dover} accepts interventions that succeed in two of three replays; CausalFlow \citep{bonagiri2026causalflow} and REFLECT \citep{lin2026reflect} verify counterfactual repairs by replay; Causal Agent Replay \citep{shah2026car} estimates intervention effects by same-policy resampling and localizes the latest significant step; minimal repair families \citep{li2026mrfr} extend replay to multi-step repairs; and Counterfactual Rollout Replay \citep{li2026crr} forks environments, substitutes an action, and continues with the same policy to train software agents. \citet{luan2026repair} show that many apparent repairs succeed by chance and must be compared against resampling, as our net-gain objective does; ReplayRCA \citep{canale2026replayrca} scores a diagnoser's single-step repair the same way, with the failed agent itself continuing. Closest to our setting, \citet{zhang2026intervention} estimate an intervention's value by prefix branching in ALFWorld with Qwen2.5-7B and Llama-3.1-8B and learn oversight controllers that transfer between the two base models with small regret; \citet{vasudev2026prediction} show that accurate failure prediction need not yield beneficial interventions. We instead select among many candidate repairs that the deployed agent itself must carry out, and measure the loss when the executor behind the evaluation differs. \citet{zhang2026credit} audits step-level credit against executed replay with the same two models without transferring labels across policies; Verified Critical Step Optimization \citep{li2026cso} keeps an expert alternative only if the policy completes the task from it; and forked rollouts show that stitching logged trajectories mis-scores model switches \citep{gonuguntla2026replaygap}. \citet{pakhomov2026executed} argue that attribution claims need executed counterfactuals; we quantify what is lost when the replay is real but not performed by the deployed policy.

\paragraph{Policy-dependent step values.}
Monte-Carlo step values depend on the completing policy, a fact used to train process reward models \citep{wang2024mathshepherd,xiong2024ipr,choudhury2025agentprm}. Such models overestimate success when a smaller model completes the trajectory, which motivates calibrating them \citep{park2025calibration}, and BayesianPRM corrects overestimation caused by a mismatch between the data-collection and test-time policies \citep{qiao2026bayesianprm}; PCCC \citep{meng2026pccc} completes counterfactuals with a frozen reference policy and reweights toward the training policy. EASE \citep{xiong2026ease} shows that skill curation tuned for one executor transfers poorly to others, and \citet{zhu2026lvrl} find that a frozen stronger LLM's value changes agree in sign with a weaker policy's Monte-Carlo rollouts only 66--72\% of the time. Pipelines that learn from failures reuse verified corrections across policy updates: ETO collects failures once and trains over several rounds \citep{song2024eto}, whereas Agent-R rebuilds its data with the current actor at every iteration \citep{yuan2025agentr}. These works calibrate or correct value estimates, or train on them; we measure what executor dependence costs when repairs are selected, including across one such update, and when it can be avoided cheaply.

\section{Limitations and Conclusion}
\label{sec:conclusion}

Our study covers three deployed agents, all open models of 2--8B parameters; four text environments, each restorable exactly to any history; and single-step repairs within fixed candidate sets. The screened pools condition on failures in which some single-step repair helps, and the unscreened pools still restrict task types and repair positions. On \tautwo{}, selection is worth too little for the pool to test mismatch. The training analysis covers one round of LoRA fine-tuning in one environment, on a candidate set that contains the planner's action; its first pair of successors was analyzed post hoc, as were the label, ranking, and loss-share analyses and the human study.

Which repair is best depends on who executes it. Evaluators that substitute for the deployed executor---greedy replay, environment oracles, LLM judges, stronger models, and the agent's own fine-tuned successor---choose repairs that cost the deployed agent real value. More rollouts of a substitute do not close the gap: twenty per candidate from Qwen2.5-14B are worth no more than two of the 7B agent's own, and on contested cases the two models prefer opposite repairs. A few rollouts of the deployed agent recover most of the loss. Repairs should be verified with the executor that will run them, and repair labels and benchmarks should state which executor, at which training version, defines them.

\section*{Impact Statement}

This paper studies how repairs for failed LLM agents are evaluated. Better-targeted repairs could make deployed agents more reliable, and our recommendation to verify repairs with the deployed agent could reduce the risk that fixes validated on a stronger model fail in deployment. We do not foresee specific negative societal consequences beyond those of improving LLM agents in general.

\bibliography{example_paper}

@InProceedings{zhang2025whowhen,
  title = 	 {Which Agent Causes Task Failures and When? {O}n Automated Failure Attribution of {LLM} Multi-Agent Systems},
  author =       {Zhang, Shaokun and Yin, Ming and Zhang, Jieyu and Liu, Jiale and Han, Zhiguang and Zhang, Jingyang and Li, Beibin and Wang, Chi and Wang, Huazheng and Chen, Yiran and Wu, Qingyun},
  booktitle = 	 {Proceedings of the 42nd International Conference on Machine Learning},
  pages = 	 {76583--76599},
  year = 	 {2025},
  editor = 	 {Singh, Aarti and Fazel, Maryam and Hsu, Daniel and Lacoste-Julien, Simon and Berkenkamp, Felix and Maharaj, Tegan and Wagstaff, Kiri and Zhu, Jerry},
  volume = 	 {267},
  series = 	 {Proceedings of Machine Learning Research},
  month = 	 {13--19 Jul},
  publisher =    {PMLR},
  url = 	 {https://proceedings.mlr.press/v267/zhang25cq.html},
  }

@misc{zhang2025agentracer,
      title={{AgenTracer}: Who Is Inducing Failure in the {LLM} Agentic Systems?}, 
      author={Guibin Zhang and Junhao Wang and Junjie Chen and Wangchunshu Zhou and Kun Wang and Shuicheng Yan},
      year={2025},
      eprint={2509.03312},
      archivePrefix={arXiv},
      primaryClass={cs.CL},
      url={https://arxiv.org/abs/2509.03312}, 
}

@inproceedings{zhu2025agentdebug,
  title={{AgentDebug}: Where {LLM} Agents Fail and How They can Learn From Failures},
  author={Kunlun Zhu and Muxin Tian and Bingxuan Li and Zijia Liu and Yingxuan Yang and Jiaxun Zhang and Pengrui Han and Qipeng Xie and Fuyang Cui and Weijia Zhang and Xiaoteng Ma and Xiaodong Yu and Gowtham Ramesh and Jialian Wu and Xiangru Tang and Zicheng Liu and Heng Ji and Pan Lu and James Zou and Jiaxuan You},
  booktitle={Workshop on Failure Modes of Agentic AI at ICML 2026},
  year={2026},
  url={https://openreview.net/forum?id=VIO4WbZuVW}
}

@misc{ma2025dover,
      title={{DoVer}: Intervention-Driven Auto Debugging for {LLM} Multi-Agent Systems}, 
      author={Ming Ma and Jue Zhang and Fangkai Yang and Yu Kang and Qingwei Lin and Saravan Rajmohan and Dongmei Zhang},
      year={2026},
      eprint={2512.06749},
      archivePrefix={arXiv},
      primaryClass={cs.AI},
      url={https://arxiv.org/abs/2512.06749}, 
}

@misc{shah2026car,
      title={Causal Agent Replay: Counterfactual Attribution for {LLM}-Agent Failures}, 
      author={Jaineet Shah},
      year={2026},
      eprint={2606.08275},
      archivePrefix={arXiv},
      primaryClass={cs.LG},
      url={https://arxiv.org/abs/2606.08275}, 
}

@misc{bonagiri2026causalflow,
      title={{CausalFlow}: Causal Attribution and Counterfactual Repair for {LLM} Agent Failures}, 
      author={Akash Bonagiri and Devang Borkar and Gerard Janno Anderias and Setareh Rafatirad and Houman Homayoun},
      year={2026},
      eprint={2605.25338},
      archivePrefix={arXiv},
      primaryClass={cs.LG},
      url={https://arxiv.org/abs/2605.25338}, 
}

@misc{zhu2026aed,
      title={Agent Error Dataset: Scaling 50,000 Error--Diagnosis Pairs for Failure Analysis and Error-Aware Post-Training}, 
      author={Kunlun Zhu and Xuyan Ye and Yibo Li and Cheng Qian and Beibin Li and Heng Ji},
      year={2026},
      eprint={2609.40111},
      archivePrefix={arXiv},
      primaryClass={cs.AI},
      url={https://arxiv.org/abs/2609.40111}, 
}

@inproceedings{xiong2024ipr,
    title = "Watch Every Step! {LLM} Agent Learning via Iterative Step-level Process Refinement",
    author = "Xiong, Weimin  and
      Song, Yifan  and
      Zhao, Xiutian  and
      Wu, Wenhao  and
      Wang, Xun  and
      Wang, Ke  and
      Li, Cheng  and
      Peng, Wei  and
      Li, Sujian",
    editor = "Al-Onaizan, Yaser  and
      Bansal, Mohit  and
      Chen, Yun-Nung",
    booktitle = "Proceedings of the 2024 Conference on Empirical Methods in Natural Language Processing",
    month = nov,
    year = "2024",
    address = "Miami, Florida, USA",
    publisher = "Association for Computational Linguistics",
    url = "https://aclanthology.org/2024.emnlp-main.93/",
    doi = "10.18653/v1/2024.emnlp-main.93",
    pages = "1556--1572",
}

@misc{choudhury2025agentprm,
      title={Process Reward Models for {LLM} Agents: Practical Framework and Directions}, 
      author={Sanjiban Choudhury},
      year={2025},
      eprint={2502.10325},
      archivePrefix={arXiv},
      primaryClass={cs.LG},
      url={https://arxiv.org/abs/2502.10325}, 
}

@misc{meng2026pccc,
      title={Policy-Conditioned Counterfactual Credit for Verifiable Reinforcement Learning of Long-Horizon Language Agents}, 
      author={Renwei Meng},
      year={2026},
      eprint={2606.05263},
      archivePrefix={arXiv},
      primaryClass={cs.LG},
      url={https://arxiv.org/abs/2606.05263}, 
}

@inproceedings{shridhar2021alfworld,
  title     = {{ALFWorld: Aligning Text and Embodied Environments for Interactive Learning}},
  author    = {Shridhar, Mohit and Yuan, Xingdi and Cote, Marc-Alexandre and Bisk, Yonatan and Trischler, Adam and Hausknecht, Matthew},
  booktitle = {International Conference on Learning Representations},
  year      = {2021},
  url       = {https://mlanthology.org/iclr/2021/shridhar2021iclr-alfworld/}
}

@misc{qwen2025qwen25,
      title={{Qwen2.5} Technical Report}, 
      author={Qwen and An Yang and Baosong Yang and Beichen Zhang and Binyuan Hui and Bo Zheng and Bowen Yu and Chengyuan Li and Dayiheng Liu and Fei Huang and Haoran Wei and Huan Lin and Jian Yang and Jianhong Tu and Jianwei Zhang and Jianxin Yang and Jiaxi Yang and Jingren Zhou and Junyang Lin and Kai Dang and Keming Lu and Keqin Bao and Kexin Yang and Le Yu and Mei Li and Mingfeng Xue and Pei Zhang and Qin Zhu and Rui Men and Runji Lin and Tianhao Li and Tianyi Tang and Tingyu Xia and Xingzhang Ren and Xuancheng Ren and Yang Fan and Yang Su and Yichang Zhang and Yu Wan and Yuqiong Liu and Zeyu Cui and Zhenru Zhang and Zihan Qiu},
      year={2025},
      eprint={2412.15115},
      archivePrefix={arXiv},
      primaryClass={cs.CL},
      url={https://arxiv.org/abs/2412.15115}, 
}

@misc{zhang2026credit,
      title={Credit Without Ground Truth: Claim-Matched Evaluation of Step-Level Credit with Finite Executed Replay}, 
      author={Haiyue Zhang},
      year={2026},
      eprint={2608.19760},
      archivePrefix={arXiv},
      primaryClass={cs.LG},
      url={https://arxiv.org/abs/2608.19760}, 
}

@misc{li2026mrfr,
      title={Localizing Emergent Failures in Agentic {AI}: Recovering Minimal Repair Families via Counterfactual Replay}, 
      author={Bingjie Li and Yumeng Song and Zhongming Yao and Tianyi Li},
      year={2026},
      eprint={2608.29228},
      archivePrefix={arXiv},
      primaryClass={cs.AI},
      url={https://arxiv.org/abs/2608.29228}, 
}

@misc{luan2026repair,
      title={Repair or Resample? Rethinking Failure Debugging in {LLM} Multi-Agent Systems}, 
      author={Zhongwen Luan and Xiaoyu Zhang and Ming Hu and Yue Yang and Jiongchi Yu and Xiaohong Chen},
      year={2026},
      eprint={2608.25920},
      archivePrefix={arXiv},
      primaryClass={cs.AI},
      url={https://arxiv.org/abs/2608.25920}, 
}

@misc{xiong2026ease,
      title={{EASE}: Behavior-Adaptive Skill Curation for Self-Evolving Agents}, 
      author={Zhen Xiong and Qiaoyu Tan},
      year={2026},
      eprint={2609.36746},
      archivePrefix={arXiv},
      primaryClass={cs.AI},
      url={https://arxiv.org/abs/2609.36746}, 
}

@inproceedings{wang2024mathshepherd,
    title = "{Math-Shepherd}: Verify and Reinforce {LLM}s Step-by-step without Human Annotations",
    author = "Wang, Peiyi  and
      Li, Lei  and
      Shao, Zhihong  and
      Xu, Runxin  and
      Dai, Damai  and
      Li, Yifei  and
      Chen, Deli  and
      Wu, Yu  and
      Sui, Zhifang",
    editor = "Ku, Lun-Wei  and
      Martins, Andre  and
      Srikumar, Vivek",
    booktitle = "Proceedings of the 62nd Annual Meeting of the {A}ssociation for {C}omputational {L}inguistics (Volume 1: Long Papers)",
    month = aug,
    year = "2024",
    address = "Bangkok, Thailand",
    publisher = "Association for Computational Linguistics",
    url = "https://aclanthology.org/2024.acl-long.510/",
    doi = "10.18653/v1/2024.acl-long.510",
    pages = "9426--9439",
}

@article{helmert2006fastdownward,
   title={The {Fast Downward} Planning System},
   volume={26},
   ISSN={1076-9757},
   url={http://dx.doi.org/10.1613/jair.1705},
   DOI={10.1613/jair.1705},
   journal={Journal of Artificial Intelligence Research},
   publisher={AI Access Foundation},
   author={Helmert, M.},
   year={2006},
   month=jul, pages={191--246} 
}

@inproceedings{kwon2023vllm,
author = {Kwon, Woosuk and Li, Zhuohan and Zhuang, Siyuan and Sheng, Ying and Zheng, Lianmin and Yu, Cody Hao and Gonzalez, Joseph and Zhang, Hao and Stoica, Ion},
title = {Efficient Memory Management for Large Language Model Serving with {PagedAttention}},
year = {2023},
isbn = {9798400702297},
publisher = {Association for Computing Machinery},
address = {New York, NY, USA},
url = {https://doi.org/10.1145/3600006.3613165},
doi = {10.1145/3600006.3613165},
booktitle = {Proceedings of the 29th Symposium on Operating Systems Principles},
pages = {611--626},
numpages = {16},
location = {Koblenz, Germany},
series = {SOSP '23}
}

@misc{liu2026whowhenpro,
      title={{Who\&When Pro}: Can {LLMs} Really Attribute Failures in {AI} Agents?}, 
      author={Jiale Liu and Huajun Xi and Shaokun Zhang and Yifan Zeng and Tianwei Yue and Chi Wang and Jian Kang and Qingyun Wu and Huazheng Wang},
      year={2026},
      eprint={2607.09996},
      archivePrefix={arXiv},
      primaryClass={cs.AI},
      url={https://arxiv.org/abs/2607.09996}, 
}

@inproceedings{qi2026trajdebug,
    title = "{TRAJDEBUG}: Tracing Error Lifecycle to Identify Critical Failures in Long-Horizon Agent Trajectories",
    author = "Qi, Yunjia  and
      Yin, Zehua  and
      Shi, Xintong  and
      Peng, Hao  and
      Lu, Songyuanyi  and
      Liu, Yixian  and
      Xuan, Richeng  and
      Liu, Yuhong  and
      Hu, Zhichao  and
      Wang, Xiaozhi  and
      Hou, Lei  and
      Xu, Bin  and
      Li, Juanzi",
    editor = "Lee, Hung-Yi  and
      Dev, Sunipa  and
      Li, Jessy  and
      Elliott, Desmond",
    booktitle = "Findings of the {A}ssociation for {C}omputational {L}inguistics: {EMNLP} 2026",
    month = oct,
    year = "2026",
    address = "Budapest, Hungary",
    publisher = "Association for Computational Linguistics",
    url = "https://aclanthology.org/2026.findings-emnlp.437/",
    pages = "9169--9192",
    ISBN = "979-8-89176-462-0",
}

@misc{xia2026agentlocate,
      title={Who Broke the System? Failure Localization in {LLM}-Based Multi-Agent Systems}, 
      author={Yufei Xia and Anjun Gao and Yueyang Quan and Zhuqing Liu and Minghong Fang},
      year={2026},
      eprint={2607.07989},
      archivePrefix={arXiv},
      primaryClass={cs.CR},
      url={https://arxiv.org/abs/2607.07989}, 
}

@misc{nguyen2026soap,
      title={Can Failed Trajectories Reveal Where Agents Go Wrong? Agentic Failure Attribution Without Step-Level Labels}, 
      author={Do Nguyen-Thanh and Junlin Fang and Bing Li and Zhen Fang and Sean Du},
      year={2026},
      eprint={2610.13105},
      archivePrefix={arXiv},
      primaryClass={cs.LG},
      url={https://arxiv.org/abs/2610.13105}, 
}

@inproceedings{li2026cso,
    title = "Verified Critical Step Optimization for {LLM} Agents",
    author = "Li, Mukai  and
      Zeng, Qingcheng  and
      Fang, Tianqing  and
      Liang, Zhenwen  and
      Song, Linfeng  and
      Liu, Qi  and
      Mi, Haitao  and
      Yu, Dong",
    editor = "Liakata, Maria  and
      Moreira, Viviane P.  and
      Zhang, Jiajun  and
      Jurgens, David",
    booktitle = "Findings of the {A}ssociation for {C}omputational {L}inguistics: {ACL} 2026",
    month = jul,
    year = "2026",
    address = "San Diego, California, United States",
    publisher = "Association for Computational Linguistics",
    url = "https://aclanthology.org/2026.findings-acl.1974/",
    doi = "10.18653/v1/2026.findings-acl.1974",
    pages = "39627--39639",
    ISBN = "979-8-89176-395-1",
}

@misc{vasudev2026prediction,
      title={Accurate Failure Prediction in Agents Does Not Imply Effective Failure Prevention}, 
      author={Rakshith Vasudev and Melisa Russak and Dan Bikel and Waseem Alshikh},
      year={2026},
      eprint={2602.03338},
      archivePrefix={arXiv},
      primaryClass={cs.CL},
      url={https://arxiv.org/abs/2602.03338}, 
}

@misc{zhu2026lvrl,
      title={{LLM} Value Functions for Credit Assignment in Agentic {RL}}, 
      author={Kaijie Zhu and Weixiang Yan and Omar Attia and Abbas Kazerouni and Alexander Toshev and Andrew Szot},
      year={2026},
      eprint={2610.15840},
      archivePrefix={arXiv},
      primaryClass={cs.LG},
      url={https://arxiv.org/abs/2610.15840}, 
}

@misc{grattafiori2024llama3,
      title={The {Llama 3} Herd of Models}, 
      author={Aaron Grattafiori and Abhimanyu Dubey and Abhinav Jauhri and Abhinav Pandey and Abhishek Kadian and Ahmad Al-Dahle and Aiesha Letman and Akhil Mathur and Alan Schelten and Alex Vaughan and others},
      year={2024},
      eprint={2407.21783},
      archivePrefix={arXiv},
      primaryClass={cs.AI},
      url={https://arxiv.org/abs/2407.21783}, 
}

@inproceedings{yao2022webshop,
 author = {Yao, Shunyu and Chen, Howard and Yang, John and Narasimhan, Karthik},
 booktitle = {Advances in Neural Information Processing Systems},
 doi = {10.52202/068431-1508},
 editor = {S. Koyejo and S. Mohamed and A. Agarwal and D. Belgrave and K. Cho and A. Oh},
 pages = {20744--20757},
 publisher = {Curran Associates, Inc.},
 title = {{WebShop}: Towards Scalable Real-World Web Interaction with Grounded Language Agents},
 url = {https://proceedings.neurips.cc/paper_files/paper/2022/file/82ad13ec01f9fe44c01cb91814fd7b8c-Paper-Conference.pdf},
 volume = {35},
 year = {2022}
}

@inproceedings{lin2021pyserini,
author = {Lin, Jimmy and Ma, Xueguang and Lin, Sheng-Chieh and Yang, Jheng-Hong and Pradeep, Ronak and Nogueira, Rodrigo},
title = {{Pyserini}: A {Python} Toolkit for Reproducible Information Retrieval Research with Sparse and Dense Representations},
year = {2021},
isbn = {9781450380379},
publisher = {Association for Computing Machinery},
address = {New York, NY, USA},
url = {https://doi.org/10.1145/3404835.3463238},
doi = {10.1145/3404835.3463238},
booktitle = {Proceedings of the 44th International ACM SIGIR Conference on Research and Development in Information Retrieval},
pages = {2356--2362},
numpages = {7},
location = {Virtual Event, Canada},
series = {SIGIR '21}
}

@inproceedings{park2025calibration,
 author = {Park, Young-Jin and Greenewald, Kristjan and Alimohammadi, Kaveh and Wang, Hao and Azizan, Navid},
 booktitle = {Advances in Neural Information Processing Systems},
 doi = {10.52202/085713-1297},
 editor = {D. Belgrave and C. Zhang and H. Lin and R. Pascanu and P. Koniusz and M. Ghassemi and N. Chen},
 pages = {38852--38895},
 publisher = {Curran Associates, Inc.},
 title = {Know What You Don\textquotesingle t Know: Uncertainty Calibration of Process Reward Models},
 url = {https://proceedings.neurips.cc/paper_files/paper/2025/file/37664246a1e07e212ddacea6e5a523f2-Paper-Conference.pdf},
 volume = {38, Main Conference},
 year = {2025}
}

@misc{zhang2026intervention,
      title={Calibration Is Not Control: Intervention Value for {LLM}-Agent Oversight}, 
      author={Chubin Zhang and Zhenglin Wan and Xingrui Yu and Jingxuan Wu and Qi Wen and Pengfei Zhou and Wangbo Zhao and Ivor Tsang},
      year={2026},
      eprint={2606.21399},
      archivePrefix={arXiv},
      primaryClass={cs.AI},
      url={https://arxiv.org/abs/2606.21399}, 
}

@misc{gonuguntla2026replaygap,
      title={The Replay Gap: Static Evaluation of Model Switching in {LLM} Agents Scores the Wrong World}, 
      author={Ashritha Gonuguntla},
      year={2026},
      eprint={2608.08239},
      archivePrefix={arXiv},
      primaryClass={cs.LG},
      url={https://arxiv.org/abs/2608.08239}, 
}

@misc{li2026crr,
      title={Counterfactual Rollout Replay: Forkable Environments as Free Process Rewards for Software Engineering Agents}, 
      author={Yuanhao Li and Hongbo Wang and Xuhong Chen and Yiming Cao and Xunzhu Tang},
      year={2026},
      eprint={2609.33875},
      archivePrefix={arXiv},
      primaryClass={cs.SE},
      url={https://arxiv.org/abs/2609.33875}, 
}

@misc{qiao2026bayesianprm,
      title={Process Reward Modeling with Reliability- and Conservatism-Aware Bayesian Belief}, 
      author={Zhongjian Qiao and Zhixiang Zhou and Jiafei Lyu and Peisong Wang and Qiaosheng Zhang and Siyang Gao and Shuang Qiu},
      year={2026},
      eprint={2610.23223},
      archivePrefix={arXiv},
      primaryClass={cs.LG},
      url={https://arxiv.org/abs/2610.23223}, 
}

@inproceedings{barke2026agentrx,
    title = "{A}gent{R}x: Diagnosing {AI} Agent Failures from Execution Trajectories",
    author = "Barke, Shraddha  and
      Goyal, Arnav  and
      Khare, Alind  and
      Singh, Avaljot  and
      Nath, Suman  and
      Bansal, Chetan",
    editor = "Lee, Hung-Yi  and
      Dev, Sunipa  and
      Li, Jessy  and
      Elliott, Desmond",
    booktitle = "Findings of the {A}ssociation for {C}omputational {L}inguistics: {EMNLP} 2026",
    month = oct,
    year = "2026",
    address = "Budapest, Hungary",
    publisher = "Association for Computational Linguistics",
    url = "https://aclanthology.org/2026.findings-emnlp.598/",
    pages = "12610--12639",
    ISBN = "979-8-89176-462-0",
}

@misc{qwen2026flash,
      title={On the Design of {Qwen3.8-Next} Architecture: Evaluation, Efficiency, and Training Stability}, 
      author={Zihan Qiu and Zekun Wang and Xiao Li and Yanpeng Li and Yang Xu and Yixuan Wang and Huaqing Zhang and Rui Men and Bochao Mao and Chengruidong Zhang and Fan Zhou and Hao Luo and Haofeng Huang and Haoran Lian and Haoyan Huang and Hongqing Chen and Jianwei Zhang and Jing Xu and Junjie Wang and Langshi Chen and Liangyu Wang and Linlang Jiang and Man Yuan and Minmin Sun and Peng Jin and Siqi Zhang and Siyu Wang and Xingzhang Ren and Yakai Wang and Yi Zhang and Yiming Dong and Yizhong Cao and Yubo Ma and Yunfei Mao and Bo Zheng and Dayiheng Liu},
      year={2026},
      eprint={2608.30320},
      archivePrefix={arXiv},
      primaryClass={cs.CL},
      url={https://arxiv.org/abs/2608.30320}, 
}

@InProceedings{barres2025tau2,
  title = 	 {{$\tau^2$-Bench}: Evaluating Conversational Agents in a Dual-Control Environment},
  author =       {Barres, Victor and Dong, Honghua and Ray, Soham and Si, Xujie and Narasimhan, Karthik R},
  booktitle = 	 {Proceedings of the 43rd International Conference on Machine Learning},
  pages = 	 {6736--6782},
  year = 	 {2026},
  editor = 	 {Zhang, Tong and Dudik, Miroslav and Jaggi, Martin and Agarwal, Alekh and Li, Sharon and Schuurmans, Dale and Zhu, Jerry and Berkenkamp, Felix and Dong, Hanze and Bietti, Alberto},
  volume = 	 {306},
  series = 	 {Proceedings of Machine Learning Research},
  month = 	 {06--11 Jul},
  publisher =    {PMLR},
  url = 	 {https://proceedings.mlr.press/v306/barres26a.html}
}

@misc{qwen2026qwen35,
  title        = {{Qwen3.5}},
  author       = {{Qwen Team}},
  year         = {2026},
  howpublished = {Model release},
  url          = {https://huggingface.co/Qwen}
}

@inproceedings{hu2022lora,
title={Lo{RA}: Low-Rank Adaptation of Large Language Models},
author={Edward J Hu and Yelong Shen and Phillip Wallis and Zeyuan Allen-Zhu and Yuanzhi Li and Shean Wang and Lu Wang and Weizhu Chen},
booktitle={International Conference on Learning Representations},
year={2022},
url={https://openreview.net/forum?id=nZeVKeeFYf9}
}

@inproceedings{song2024eto,
    title = "Trial and Error: Exploration-Based Trajectory Optimization of {LLM} Agents",
    author = "Song, Yifan  and
      Yin, Da  and
      Yue, Xiang  and
      Huang, Jie  and
      Li, Sujian  and
      Lin, Bill Yuchen",
    editor = "Ku, Lun-Wei  and
      Martins, Andre  and
      Srikumar, Vivek",
    booktitle = "Proceedings of the 62nd Annual Meeting of the {A}ssociation for {C}omputational {L}inguistics (Volume 1: Long Papers)",
    month = aug,
    year = "2024",
    address = "Bangkok, Thailand",
    publisher = "Association for Computational Linguistics",
    url = "https://aclanthology.org/2024.acl-long.409/",
    doi = "10.18653/v1/2024.acl-long.409",
    pages = "7584--7600"
}

@misc{yuan2025agentr,
      title={{Agent-R}: Training Language Model Agents to Reflect via Iterative Self-Training}, 
      author={Siyu Yuan and Zehui Chen and Zhiheng Xi and Junjie Ye and Zhengyin Du and Jiecao Chen},
      year={2025},
      eprint={2501.11425},
      archivePrefix={arXiv},
      primaryClass={cs.AI},
      url={https://arxiv.org/abs/2501.11425}, 
}

@misc{pakhomov2026executed,
      title={Attributing Agent Failures to Their History Needs an Executed Counterfactual: Toward a Reporting Standard}, 
      author={Egor Pakhomov and Erik Nijkamp},
      year={2026},
      eprint={2611.05650},
      archivePrefix={arXiv},
      primaryClass={cs.AI},
      url={https://arxiv.org/abs/2611.05650}, 
}

@misc{lin2026reflect,
      title={{REFLECT}: Intervention-Supported Error Attribution for Silent Failures in {LLM} Agent Traces}, 
      author={Xiaofeng Lin and Yingxu Wang and Tung Sum Thomas Kwok and Daniel Guo and Sahil Arun Nale and Charles Fleming and Guang Cheng},
      year={2026},
      eprint={2606.09071},
      archivePrefix={arXiv},
      primaryClass={cs.AI},
      url={https://arxiv.org/abs/2606.09071}, 
}

@misc{canale2026replayrca,
      title={{ReplayRCA}: Measuring Post-hoc Diagnosis and Live Oversight of Agent Failures by Counterfactual Replay}, 
      author={Jeremy Canale and Fabien Soulis},
      year={2026},
      eprint={2610.41025},
      archivePrefix={arXiv},
      primaryClass={cs.AI},
      url={https://arxiv.org/abs/2610.41025}, 
}
\bibliographystyle{icml2026}

\newpage
\appendix

\section{Analysis Plan, Status, and Statistics}
\label{app:plan}

\subsection{Status of Every Analysis}
\label{app:audit}

All designs, tests, thresholds, and predictions were written into an analysis plan that we maintained as a dated internal record: each entry records its date and time and what had and had not been seen when it was written, and later amendments and corrections were added as dated notes. The plan was not independently timestamped or publicly registered; we release it, in anonymized form, together with the code. \Cref{tab:audit} summarizes, for each analysis in the paper, whether it was specified before its data were collected, whether it used fresh failures or fresh rollouts, and how we report it. ``Pre-specified'', ``fresh data'', ``fresh rollouts'', and ``exploratory'' are separate dimensions: for example, the reversal checks were pre-specified and used only fresh rollouts, but chose their cases with earlier rollouts.

\begin{table*}[t]
  \caption{Status of every analysis. Pre-specified: written into the dated plan before the data it uses were collected or computed. Fresh: new failures (F) or new rollouts on existing branches (R). Confirmatory tests are Holm-corrected within the family shown; ``secondary'' tests use a lower bound above 0 without correction.}
  \label{tab:audit}
  \centering
  \footnotesize
  \setlength{\tabcolsep}{4pt}
  \begin{tabular}{@{}>{\raggedright\arraybackslash}p{0.27\textwidth}>{\raggedright\arraybackslash}p{0.2\textwidth}>{\raggedright\arraybackslash}p{0.2\textwidth}c>{\raggedright\arraybackslash}p{0.22\textwidth}@{}}
    \toprule
    Analysis & Pool & Pre-specified & Fresh & Reported as \\
    \midrule
    C1--C3; C4--C6 & screened gridworld & yes & F & confirmatory (Holm, 3); secondary \\
    Greedy with escape (E1, E2) & screened gridworld & yes, before running & R & secondary \\
    B2--B4; B5--B9, informed judge & screened ALFWorld & yes & F & confirmatory (Holm, 3); secondary \\
    Other executors X1, X2 (14B, flash) & both screened & yes, before running & R & secondary \\
    Published decision rules & both screened & no & --- & exploratory \\
    Labels and rankings (\cref{sec:consequences}) & both screened & no & --- & exploratory \\
    Privileged screen & both screened & fixed on gridworld & --- & exploratory on ALFWorld \\
    P1, P2; S1, S2; budget curves; screens & unscreened ALFWorld (7B) & yes & F & confirmatory (Holm, 2); secondary; descriptive \\
    H1--H3 & unscreened ALFWorld (both) & yes & F, R & confirmatory (Holm, 3) \\
    LLM-proposed subset & unscreened ALFWorld & after full results & --- & exploratory \\
    Saturation strata; cross-pool transfers & unscreened ALFWorld & yes, before computing & --- & pre-specified descriptive \\
    Reversal checks D$_{14}$, D$_{8}$ & unscreened ALFWorld (7B) & yes (D$_8$ by trigger) & R & secondary; cases chosen with earlier rollouts \\
    LLM-located pipeline & unscreened ALFWorld (7B) & yes, before running & R & exploratory follow-up \\
    P1w, P2w, H1w, H2w, H3w & unscreened WebShop & yes, before any formal run & F & confirmatory (Holm, 5) \\
    Human reader & unscreened ALFWorld (7B) & readouts drafted; protocol not frozen & --- & exploratory \\
    Unscreened gridworld & 100 gridworld failures & yes & F & secondary \\
    Shared-repair cost $G$ (\cref{app:share}) & both screened & yes & --- & inconclusive; not used \\
    P1$\tau$, P2$\tau$; sensitivity analyses & unscreened \tautwo{} & yes, before the 2B rollouts & F & positive control not met; reported as bounds \\
    Loss as a share of selection value & all unscreened & after full results & --- & exploratory \\
    Per-position labels (\cref{app:labels}) & all unscreened & after full results & --- & exploratory \\
    Per-position staleness of labels & unscreened ALFWorld (7B); \tautwo{} & yes, in a separate dated plan & R & descriptive here \\
    Successor as evaluator, first pair (\cref{sec:training}) & unscreened ALFWorld (7B) & after the successors' rollouts & --- & exploratory \\
    Successor as evaluator, second pair (N1) & unscreened ALFWorld (7B) & yes, before training & R & confirmed for both seeds \\
    \bottomrule
  \end{tabular}
\end{table*}

\subsection{Statistical Details}
\label{app:stats}

For every evaluator and trajectory we compute the deployment net gain of the selected repair from deployed-agent rollouts that the selection did not use. Ties in any selection rule are broken uniformly at random in expectation. Failed API calls of Qwen3.8-flash are retried and never counted as policy failures. One-sided $p$-values are $(\#\{\text{replicates}\le0\}+1)/(B+1)$ with $B=2{,}000$.

\paragraph{Unscreened pools.} The bootstrap resamples trajectories with replacement and recomputes every selection and evaluation (nonparametric percentile intervals). Matched selection selects with five of each branch's ten rollouts and evaluates with the other five, averaged over 50 random splits. Random numbers come from one seeded stream per trajectory and executor, so that adding an executor leaves the other readouts unchanged; the analysis that produced P1 and P2 used a single shared stream (P1 differs by 0.001 between the two).

\paragraph{Screened pools.} Each bootstrap replicate resamples trajectories with replacement and, for every stochastic branch, redraws its success count from a binomial distribution with success probability $(k+0.5)/(n+1)$, where $k$ of $n$ rollouts succeeded; all selections and evaluations are recomputed in every replicate. The shrinkage keeps resampled probabilities away from 0 and 1, but it also makes the true values of the bootstrap world systematically lower than the point estimates when many branches have extreme success rates, which shifts the whole bootstrap distribution downward (by 0.01--0.03 in the gridworld and 0.03--0.07 in ALFWorld). Because cross-fitted estimates are unbiased within the bootstrap world, we re-center the bootstrap distribution on the point estimate. This correction was added after the pre-specified analyses had been run; the uncentered percentile intervals lead to the same decisions for every pre-specified test (\cref{tab:registered}).

\paragraph{H1: recorded and regenerated values.} The pre-specified analysis of H1 gave 0.074 [0.038, 0.113], which we report. Afterwards the Llama rollouts of 65 of the 1,286 branches of the 7B ALFWorld pool were lost and regenerated with the same seeds and code; because vLLM sampling is not bitwise reproducible, the released data give 0.071 [0.037, 0.108], with the same decision. All other numbers that use these rollouts (the cross-pool comparisons, the Llama saturation stratum, the trigger of the 7B--Llama reversal check, and the post-repair H1) come from the regenerated data.

\paragraph{Decision rules.} Primary tests are Holm-corrected within the families of \cref{tab:audit}. Three secondary tests use a pre-specified pair of rules against the margin $\delta=0.10$: a loss is \emph{detected} if the lower bound exceeds 0, and \emph{within margin} if the upper bound is below $\delta$. The two are not mutually exclusive, so we report them separately; no test in the paper met both.

\begin{table*}[p]
  \caption{All pre-specified tests. Estimates are differences in deployment net gain (first minus second; reward in WebShop). Screened pools: pre-specified (uncentered) and bias-corrected 90\% intervals. Unscreened pools: nonparametric 90\% intervals. Holm: primary test, Holm-corrected within its family at one-sided $\alpha=0.05$ (screened pools: largest $p$-value 0.050 for B4 under the pre-specified bootstrap, 0.004 after correction; unscreened: P1--P2, H1--H3, and the five WebShop tests form three families, largest $p$-value 0.0005). Detected: lower bound $>0$; within margin: upper bound $<\delta=0.10$. Every decision is the same under both intervals.}
  \label{tab:registered}
  \centering
  \footnotesize
  \setlength{\tabcolsep}{3pt}
  \resizebox{\textwidth}{!}{%
  \begin{tabular}{@{}llcccll@{}}
    \toprule
    & Contrast & Estimate & Pre-specified & Bias-corrected & Rule & Decision \\
    \midrule
    \multicolumn{7}{@{}l}{\emph{Screened gridworld (34 trajectories)}} \\
    PC & matched $-$ random & 0.656 & [0.548, 0.709] & [0.574, 0.735] & lower $>0.10$ & pass \\
    C1 & matched $-$ random solvable action, irreversible point & 0.308 & [0.193, 0.371] & [0.218, 0.396] & Holm & confirmed \\
    C2 & matched $-$ greedy replay & 0.225 & [0.136, 0.284] & [0.152, 0.300] & Holm & confirmed \\
    C3 & matched $-$ solvability oracle & 0.186 & [0.090, 0.235] & [0.115, 0.261] & Holm & confirmed \\
    C4 & matched $-$ 3 rollouts & 0.109 & [0.044, 0.145] & [0.060, 0.161] & lower $>0$ & detected \\
    C5 & oracle with slack $-$ oracle & 0.149 & [0.092, 0.195] & [0.099, 0.202] & lower $>0$ & detected \\
    C6 & matched $-$ temperature 0.3 & 0.027 & [$-$0.026, 0.096] & [$-$0.030, 0.093] & both & within margin; not detected \\
    E1 & greedy with escape (10) $-$ greedy & 0.101 & [0.032, 0.169] & [0.033, 0.171] & lower $>0$ & detected \\
    E2 & deployed (10) $-$ greedy with escape (10) & 0.118 & [0.055, 0.183] & [0.057, 0.185] & within $\pm0.10$ & not within \\
    X1 & deployed (10) $-$ 14B (10) & 0.061 & [0.010, 0.098] & [0.017, 0.105] & lower $>0$ & detected \\
    X2 & deployed (10) $-$ Qwen3.8-flash (10) & 0.163 & [0.085, 0.219] & [0.098, 0.232] & lower $>0$ & detected \\
    \midrule
    \multicolumn{7}{@{}l}{\emph{Screened ALFWorld (30 trajectories)}} \\
    PC & matched $-$ random & 0.753 & [0.619, 0.740] & [0.689, 0.811] & lower $>0.10$ & pass \\
    B2 & matched $-$ greedy replay & 0.177 & [0.068, 0.217] & [0.103, 0.253] & Holm & confirmed \\
    B3 & matched $-$ solvability oracle$^\dagger$ & 0.753 & [0.619, 0.740] & [0.689, 0.811] & Holm & confirmed \\
    B4 & matched $-$ 3 rollouts & 0.070 & [0.000, 0.085] & [0.027, 0.113] & Holm & confirmed \\
    B5 & oracle with slack $-$ oracle & 0.477 & [0.376, 0.520] & [0.405, 0.549] & lower $>0$ & detected \\
    B6 & matched $-$ oracle with slack & 0.276 & [0.155, 0.315] & [0.197, 0.356] & both & detected; not within margin \\
    B7 & matched $-$ planner action (random position) & 0.253 & [0.144, 0.281] & [0.184, 0.321] & lower $>0$ & detected \\
    B8 & matched $-$ reasoning LLM judge & 0.531 & [0.352, 0.598] & [0.408, 0.655] & both & detected; not within margin \\
    B9 & matched $-$ deployed model as judge & 0.736 & [0.566, 0.760] & [0.636, 0.830] & lower $>0$ & detected \\
    J1 & matched $-$ informed LLM judge & 0.186 & [0.066, 0.245] & [0.100, 0.280] & lower $>0$ & detected \\
    X1 & deployed (10) $-$ 14B (10) & 0.291 & [0.170, 0.376] & [0.191, 0.397] & lower $>0$ & detected \\
    X2 & deployed (10) $-$ Qwen3.8-flash (10) & 0.628 & [0.511, 0.664] & [0.550, 0.704] & lower $>0$ & detected \\
    \midrule
    \multicolumn{7}{@{}l}{\emph{Unscreened ALFWorld, deployed Qwen2.5-7B (60 trajectories)}} \\
    PC & matched $-$ random & 0.707 & [0.652, 0.758] & --- & lower $>0.10$ & pass \\
    P1 & matched $-$ greedy replay & 0.135 & [0.090, 0.183] & --- & Holm & confirmed \\
    P2 & deployed (5) $-$ 14B (5) & 0.255 & [0.193, 0.316] & --- & Holm & confirmed \\
    S1 & matched $-$ planner action (earliest position) & 0.140 & [0.075, 0.209] & --- & lower $>0$ & detected \\
    S2 & matched $-$ oracle with slack & 0.195 & [0.141, 0.251] & --- & lower $>0$ & detected \\
    H1 & deployed (5) $-$ Llama-3.1-8B (5) & 0.074 & [0.038, 0.113] & --- & Holm & confirmed \\
    D$_{14}$ & reversal check, 7B--14B, 37 contested trajectories & 0.509 & [0.382, 0.641] & --- & lower $>0$ & detected \\
    D$_{8}$ & reversal check, 7B--Llama, 34 contested trajectories$^\S$ & 0.202 & [0.122, 0.287] & --- & lower $>0$ & detected$^\S$ \\
    \midrule
    \multicolumn{7}{@{}l}{\emph{Unscreened ALFWorld, deployed Llama-3.1-8B (60 trajectories)}} \\
    PC & matched $-$ random & 0.737 & [0.708, 0.763] & --- & lower $>0.10$ & pass \\
    H2 & deployed (5) $-$ Qwen2.5-7B (5) & 0.098 & [0.059, 0.141] & --- & Holm & confirmed \\
    H3 & matched $-$ greedy replay & 0.082 & [0.048, 0.122] & --- & Holm & confirmed \\
    S1 & matched $-$ planner action (earliest position) & 0.199 & [0.134, 0.266] & --- & lower $>0$ & detected \\
    S2 & matched $-$ oracle with slack & 0.217 & [0.165, 0.273] & --- & lower $>0$ & detected \\
    \midrule
    \multicolumn{7}{@{}l}{\emph{Unscreened WebShop, deployed Qwen2.5-7B (60 trajectories; reward)}} \\
    PC & matched $-$ random & 0.163 & [0.128, 0.200] & --- & lower $>0.10$ & pass \\
    P1w & matched $-$ greedy replay & 0.061 & [0.036, 0.086] & --- & Holm & confirmed \\
    H1w & deployed (5) $-$ Llama-3.1-8B (5) & 0.081 & [0.046, 0.114] & --- & Holm & confirmed \\
    P2w & deployed (5) $-$ 14B (5) & 0.073 & [0.046, 0.101] & --- & Holm & confirmed \\
    S1w & matched $-$ privileged action (earliest position) & 0.144 & [0.100, 0.190] & --- & lower $>0$ & detected \\
    \midrule
    \multicolumn{7}{@{}l}{\emph{Unscreened WebShop, deployed Llama-3.1-8B (60 trajectories; reward)}} \\
    PC & matched $-$ random & 0.314 & [0.258, 0.373] & --- & lower $>0.10$ & pass \\
    H2w & deployed (5) $-$ Qwen2.5-7B (5) & 0.166 & [0.117, 0.217] & --- & Holm & confirmed \\
    H3w & matched $-$ greedy replay & 0.131 & [0.083, 0.181] & --- & Holm & confirmed \\
    S1w & matched $-$ privileged action (earliest position) & 0.248 & [0.190, 0.305] & --- & lower $>0$ & detected \\
    \midrule
    \multicolumn{7}{@{}l}{\emph{Unscreened \tautwo{}, deployed Qwen3.5-2B (115 conversations, 96 tasks)}} \\
    PC & matched $-$ random & 0.099 & [0.067, 0.133] & --- & lower $>0.10$ & not met \\
    P1$\tau$ & matched $-$ greedy replay & 0.016 & [0.000, 0.033] & --- & Holm & uninformative (PC) \\
    P2$\tau$ & deployed (5) $-$ Qwen3.5-4B (5) & 0.026 & [0.005, 0.048] & --- & Holm & uninformative (PC) \\
    \midrule
    \multicolumn{7}{@{}l}{\emph{Unscreened ALFWorld, deployed Qwen2.5-7B; selector: successor trained without the planner's candidates}} \\
    N2 & successor changed: own success after $-$ before, seeds 1 and 2 & 0.193, 0.094 & [0.132, 0.256], [0.036, 0.152] & --- & lower $>0$ & detected \\
    N1 & deployed (5) $-$ successor (5), seeds 1 and 2 & 0.557, 0.548 & [0.483, 0.634], [0.476, 0.623] & --- & lower $>0.10$ & confirmed \\
    \bottomrule
  \end{tabular}}

  \vspace{2pt}
  {\footnotesize $^\dagger$Every screened ALFWorld candidate leaves the task solvable, so the oracle selects at random and B3 coincides with the positive control; it carries no independent information. D: difference of preferences $\Delta_{\mathrm{7B}}-\Delta_{X}$ on 30 fresh rollouts per branch from each executor (\cref{tab:reversal}); contested trajectories are those in which the two executors' best candidates differed in the earlier rollouts. Both D checks are on the 7B pool. $^\S$Run because its pre-specified trigger was met (deployed 7B with 5 rollouts $-$ Llama with 10: 0.044 [0.009, 0.082]); the difference comes almost entirely from the 7B side (\cref{tab:reversal}). H1: value of the pre-specified analysis; the released data give 0.071 [0.037, 0.108] (\cref{app:stats}).}
\end{table*}

\subsection{Pre-Specified Shared-Repair Analysis}
\label{app:share}

This analysis does not carry any argument in the main text; we report it because it was pre-specified. Before computing it, we specified an analysis of the shared-repair cost $G=\frac12\sum_{e}[R_e(\hat r_e)-R_e(\hat r_{\mathrm{shared}})]$ for the 7B and 14B executors, where $\hat r_e$ is executor $e$'s own choice and $\hat r_{\mathrm{shared}}$ maximizes the average of the two executors' estimates, together with a decision rule. In ALFWorld, $G$ exceeds its value under a null in which both executors share the pooled success rates by 0.07 [0.04, 0.10]; in the gridworld by 0.02 [0.01, 0.04]. Because the null-world bias of the estimator exceeded the pre-specified tolerance of 0.05 in the gridworld (0.056), the pre-specified rule classifies the result as inconclusive. A second reference added after seeing the data, which aligned one executor's ranking with the other's, turned out to be invalid because clipping to $[0,1]$ changed the reassigned values in about half of the trajectories; we do not use it. Simulations also show that $G$ is inflated when one executor is nearly indifferent among candidates, as Qwen3.8-flash is in ALFWorld, so we report $G$ for the 7B--14B pair only.

\section{Environments, Agents, and Pools}
\label{app:setup}

\paragraph{Environments.}
In the gridworld, ALFWorld, and WebShop, an episode can be restored exactly to any intermediate history by replaying its action prefix (verified across instances and processes); \tautwo{} is described at the end of this section. The \emph{gridworld} is a fully observed $5\times5$ world: the agent must pick up a key, open a locked door, pick up the gem behind it, and return to the start within 30 steps, and a trap ends the episode when stepped on. We compute exactly whether the task is still solvable from any history ($F(h)\in\{0,1\}$) and how many steps of slack remain; stepping on the trap or wasting too many steps creates an \emph{irreversible point}, after which no action can succeed. \emph{ALFWorld} \citep{shridhar2021alfworld} is a partially observed household environment with a 50-step budget, in which the contents of closed containers are unknown until opened. It has no irreversible actions, so every failure is an exhausted budget. A Fast Downward planner \citep{helmert2006fastdownward} with access to the full state gives a plan length $L(h)$, from which we define solvability $F(h)=\mathbf{1}[L(h)\le T-t]$ and slack $T-t-L(h)$; we exclude the two-object task type, on which the 7B agent almost never succeeds. \emph{WebShop} \citep{yao2022webshop} is a simulated online store: the agent searches, opens products, selects options, and buys, and the purchase ends the episode with a reward in $[0,1]$ that scores it against the instruction. We use the original code with its 100,000-product configuration, the 1,021 human-written instructions on those products, BM25 search \citep{lin2021pyserini}, and a 25-step budget. Because rollouts rarely reach full reward, the primary objective in WebShop, fixed before any formal run, is reward; success (reward 1) is a secondary readout (\cref{app:webshop}).

\paragraph{Agents.}
The deployed agents, whose failures we repair and whose success defines deployment value, are Qwen2.5-7B-Instruct \citep{qwen2025qwen25} and, on the unscreened pools, Llama-3.1-8B-Instruct \citep{grattafiori2024llama3}, a model of similar size from another family (on \tautwo{}, Qwen3.5-2B; see below). Both are served with vLLM \citep{kwon2023vllm} at temperature 0.7 and top-$p$ 0.95, with a separate random seed for every request. At each step the agent writes a short thought followed by one action. The prompt contains the task, the current observation, the admissible actions (ALFWorld and WebShop), and the history of actions and environment feedback; earlier thoughts are not kept, so a substituted action never contradicts stale reasoning, and an inadmissible answer is re-asked up to twice without consuming a step. Other executors---Qwen2.5-14B-Instruct, Qwen3.8-flash \citep{qwen2026flash} (API, without reasoning), and each deployed model on the other's pool---use exactly the same prompt, history protocol, decoding parameters, and action parsing. The LLM judge is Qwen3.8-flash with reasoning: it reads the trajectory and the candidate list, is told which model will execute the repair, and names one candidate.

\paragraph{Pools.}
Each \emph{unscreened} pool takes the first 60 failures of fresh episodes in collection order without looking at any repair signal (in WebShop, a failure is a reward below 1). ``Unscreened'' means not screened by estimated repair value: the pools still exclude ALFWorld's two-object task, and repairs are placed only at positions that leave room to recover (at least five steps of planner slack in ALFWorld, at least five remaining steps in WebShop); no failure lacked such a position. The \emph{screened} pools, collected first, keep only failures in which some single-step repair helps, according to screening rollouts that are never reused: in the gridworld, 34 of the 107 failures by the trap or by timeout on the way back with the gem, in which some alternative at the irreversible point succeeded in at least 3 of 10 rollouts; in ALFWorld, 30 drawn at random from the 78 of 94 failures in which some alternative beat the original action by at least two successes in four rollouts each. We use the screened pools as a controlled setting in which to compare many evaluators and to analyze mechanisms.

\paragraph{Positions and candidates.}
Positions are fixed before candidates are generated. In the gridworld we intervene at the irreversible point and at three earlier solvable positions, with all six legal alternatives as candidates. In ALFWorld and WebShop, eligible positions are split into early, middle, and late thirds and one position is drawn from each. Screened ALFWorld uses the planner's next action plus random admissible commands. In the unscreened pools, each position's candidates are the union of two alternatives from each of three LLMs that read the task, the full failed log, and the admissible actions (the deployed model, Qwen3.8-flash, and Qwen2.5-14B on the 7B ALFWorld pool or else the other deployed model), a privileged action computed from the hidden state (ALFWorld: the planner's next action; WebShop: the next action toward the target product and its options), and one random admissible action; in WebShop, whose pages offer a bounded set of clickable elements, we also add every clickable alternative. Privileged and exhaustive candidates keep the candidate sets from reflecting any single proposer's preferences.

\paragraph{\tautwo{}.}
We use the retail and airline domains, with the 74 retail tasks whose outcome is scored deterministically and all 50 airline tasks. The simulated user is Qwen3.8-27B (greedy, thinking off) and is the same for every executor; the agents are Qwen3.5 models with thinking off. When a rollout continues from a prefix, the default evaluator replays the prefix's writes into the reference database and can score an inherited error as correct, so we re-score every rollout against the original task. The 2B agent ran 20 conversations per task, and up to four failures per task were kept. Positions lie before the first erroneous write; candidates are the original action, up to three distinct proposals from eight samples of Qwen3.5-4B, and the environment's reference next action. A candidate is kept only if a deployed agent could produce it: every argument must be recoverable from the conversation, the domain policy, or the tool schema, a write must have been confirmed by the user, and in retail the user must have been authenticated where the policy requires it. The 4B rollouts (six per branch) were collected on a larger pool of 800 positions for the training analysis of \cref{app:training}; the subset used here (115 failed conversations, 200 positions, 727 branches, 96 tasks) was fixed by a hash before any 4B outcome was seen.

\subsection{WebShop}
\label{app:webshop}

\paragraph{Environment.}
We use the original WebShop code and text environment \citep{yao2022webshop} without modification, with the 100,000-product configuration that the code supports (the first 100,000 products of the full product file and their attributes) and the 1,021 human-written instructions on those products; search uses the original BM25 index built with Pyserini 2.3.0 \citep{lin2021pyserini}. Prices and price limits, which the original code draws at random on start-up, are fixed by seeding. The admissible actions are those the environment lists (the search box on the search page and every clickable element); an action outside the list does nothing and consumes a step. The observation is the rich-text page; the prompt keeps the current page up to 6,000 characters and the first 300 characters of each earlier page. An episode ends with a purchase or after 25 steps. We fixed the 25-step budget before any formal run, after a 15-step smoke test in which Llama ran out of steps in 80\% of episodes; with 25 steps, the 7B agent succeeded in 23\% and Llama in 22\% of 60 smoke episodes. Replaying an action prefix reproduces every page exactly across processes, which we verified on random walks and on all 180 formal positions.

\paragraph{Pool and candidates.}
The 7B agent ran one episode on each of the first 100 instructions of a fixed shuffled order (disjoint from the smoke tests) and succeeded in 28; we took the first 60 of the 72 failures, with none excluded. Positions are drawn from steps that leave at least five steps (early, middle, and late thirds). At each position, the candidates are every clickable alternative on the page; two alternatives each from Qwen2.5-7B, Llama-3.1-8B (temperature 0), and Qwen3.8-flash (without reasoning), which read the instruction, the failed trajectory and its outcome, the current page and its admissible actions, and are told which model will execute the repair (on the search page they write their own query); a privileged action computed from the target product (search for its title; on a results page, click it if listed and otherwise go back; on its page, select the next missing target option, or buy once all are selected; on any other product page, go back); and one random action (on the search page, a query of two or three words drawn from the instruction). This gives 2,251 branches (63 with a determined outcome, such as purchases), 1--50 per position (mean 11.5). Proposal validity was 90\% (7B), 88\% (Llama), and 91\% (flash). The Llama pool was built the same way from a disjoint set of instructions: Llama succeeded in 14 of 100 episodes, and we took the first 60 of its 86 failures (none excluded), giving 1,999 branches (71 determined), 1--103 per position (mean 10.1); proposal validity was 87\% (7B), 76\% (Llama), and 85\% (flash).

\paragraph{Objective.}
The value of a branch is the mean reward of its rollouts; a purchase branch takes its single reward. In a 30-instruction test of the full pipeline before the formal runs, only 7\% of rollouts reached full reward, so we changed the primary objective from success to reward before any formal run and kept success as a pre-specified secondary readout. On all stochastic branches of the 7B pool, the 7B agent reaches full reward in 12\% of rollouts (mean reward 0.37) and Llama in 6\% (0.18); on the Llama pool, Llama reaches it in 8\% (0.20) and the 7B agent in 21\% (0.42).

\section{Unscreened Pools: Further Results}
\label{app:unscreened}

\subsection{Budget Curves and Reversal Checks}
\label{app:budget}

\begin{table}[!htbp]
  \caption{Budget curves on the unscreened pools: deployment net gain of the repair selected with $k$ rollouts per candidate from each executor, scored on the deployed agent (G: greedy replay). ALFWorld: success probability; WebShop: reward.}
  \label{tab:budgetcurve}
  \centering
  \footnotesize
  \setlength{\tabcolsep}{3pt}
  \begin{tabular}{@{}lccc|cc@{}}
    \toprule
    & \multicolumn{3}{c|}{ALFWorld, 7B pool} & \multicolumn{2}{c}{ALFWorld, Llama pool} \\
    $k$ & 7B & 14B & Llama & Llama & 7B \\
    \midrule
    1 & 0.63 & 0.38 & 0.55 & 0.68 & 0.57 \\
    2 & 0.77 & 0.49 & 0.67 & 0.82 & 0.68 \\
    3 & 0.80 & 0.53 & 0.72 & 0.85 & 0.74 \\
    5 & 0.82 & 0.57 & 0.75 & 0.87 & 0.77 \\
    10 & --- & 0.60 & 0.78 & --- & 0.79 \\
    20 & --- & 0.63 & --- & --- & --- \\
    G & 0.69 & 0.45 & 0.62 & 0.79 & 0.64 \\
    \bottomrule
  \end{tabular}

  \vspace{6pt}
  \begin{tabular}{@{}lccc|cc@{}}
    \toprule
    & \multicolumn{3}{c|}{WebShop, 7B pool} & \multicolumn{2}{c}{WebShop, Llama pool} \\
    $k$ & 7B & 14B & Llama & Llama & 7B \\
    \midrule
    1 & 0.072 & 0.042 & 0.026 & 0.095 & 0.021 \\
    2 & 0.106 & 0.052 & 0.047 & 0.209 & 0.063 \\
    3 & 0.125 & 0.066 & 0.059 & 0.263 & 0.091 \\
    5 & 0.146 & 0.073 & 0.066 & 0.310 & 0.144 \\
    10 & --- & 0.081 & 0.080 & --- & 0.201 \\
    20 & --- & 0.091 & --- & --- & --- \\
    G & 0.086 & 0.053 & 0.076 & 0.179 & 0.031 \\
    \bottomrule
  \end{tabular}

  \vspace{2pt}
  {\footnotesize 90\% intervals (nonparametric): ALFWorld 7B pool, 7B $k=5$ [0.76, 0.88], 14B $k=20$ [0.56, 0.71]; WebShop 7B pool, 7B $k=5$ [0.109, 0.187], 14B $k=20$ [0.046, 0.138], Llama $k=10$ [0.038, 0.123]; WebShop Llama pool, Llama $k=5$ [0.248, 0.373], 7B $k=10$ [0.133, 0.272].}
\end{table}

\paragraph{Budget curves.}
\Cref{tab:budgetcurve} lists the full curves. On the 7B ALFWorld pool, the 14B model's selections improve from 0.38 with one rollout per candidate to 0.63 with twenty, the value of a single deployed-agent rollout per candidate; per rollout, the 14B model costs 1.6--2.0 times the GPU time of the 7B model on the same A800 GPU. On the Llama pool, ten 7B rollouts per candidate select worse repairs for Llama (0.79) than two of Llama's own (0.82), and about as well as Llama's greedy replay. In WebShop, a Llama rollout costs about twice a 7B rollout on the same GPU (1.18 versus 0.56 seconds on the 7B pool, 1.00 versus 0.51 on the Llama pool), whereas a 14B rollout costs about the same (0.60 seconds), because the 14B model ends its episodes sooner; twenty 14B rollouts per candidate select repairs for the 7B agent worth about its own greedy replay (0.091 versus 0.086). On the WebShop Llama pool, the 7B model's greedy replay selects almost no value for Llama (0.031), against 0.179 for Llama's own. The ALFWorld 7B-pool curves come from the analysis that also computes the saturation strata; they differ from the primary analysis by at most 0.005 because the two analyses use different random streams for cross-fitting.

\begin{table}[!htbp]
  \caption{Reversal checks on the unscreened 7B pool (only fresh rollouts; 30 per branch and executor). $\Delta_e=R_e(r_a)-R_e(r_b)$ with $R_e$ the net gain of \cref{eq:netgain}, where $r_a$ is the deployed 7B agent's best candidate and $r_b$ the other executor's in the earlier rollouts; difference $=\Delta_{\mathrm{7B}}-\Delta_{\mathrm{other}}$. Control: trajectories where both executors' best candidate was $r_a$, with $r_b$ the 7B agent's second best. Sig.\ opposite / aligned: both $\Delta$s have 90\% intervals excluding 0, with opposite / equal signs.}
  \label{tab:reversal}
  \centering
  \footnotesize
  \setlength{\tabcolsep}{3pt}
  \resizebox{\columnwidth}{!}{%
  \begin{tabular}{@{}lccc@{}}
    \toprule
    & 14B, contested & 14B, control & Llama, contested \\
    \midrule
    Trajectories & 37 & 15 & 34 \\
    $\Delta_{\mathrm{7B}}$ & 0.33 [0.23, 0.44] & 0.13 [0.05, 0.23] & 0.18 [0.10, 0.27] \\
    $\Delta_{\mathrm{other}}$ & $-$0.18 [$-$0.25, $-$0.11] & 0.12 [0.05, 0.20] & $-$0.02 [$-$0.07, 0.03] \\
    Difference & 0.51 [0.38, 0.64] & 0.01 [$-$0.08, 0.11] & 0.20 [0.12, 0.29] \\
    Sig.\ opposite & 8 & 0 & 2 \\
    Sig.\ aligned & 0 & 3 & 3 \\
    \bottomrule
  \end{tabular}}
\end{table}

\paragraph{Reversal checks.}
\Cref{tab:reversal} gives the details. For the 7B--14B pair, the same contested trajectories show a difference of 0.65 in the earlier rollouts that selected them, so the fresh rollouts retain about 78\% of it; the 14B model's preference retains 83\%. For the 7B--Llama pair, the 7B agent's preference retains 83\% (0.22 to 0.18) while Llama's shrinks from $-$0.14 to $-$0.02, mostly selection noise. Under the post-repair objective, the 7B--14B difference is 0.33 [0.23, 0.44].

\subsection{WebShop Results}
\label{app:webshopres}

\Cref{tab:webshop} lists the readouts. All five primary tests are confirmed (Holm with $m=5$), and the success readouts are weaker and wider, as predicted. Matched selection gains 0.146 [0.109, 0.187] in reward on the 7B pool and 0.310 [0.248, 0.373] on the Llama pool; 85\% and 90\% of trajectories have a candidate whose estimated net reward gain exceeds 0.10 (50\% under success in both pools). Exploratory descriptions: the mean net reward gain of candidates from each source is close to zero or negative for the 7B agent and Llama (for the 7B agent, from $-$0.04 for Llama's proposals to $+$0.03 for the privileged action), and candidates proposed by an executor are not better for that executor than for the other (interaction 0.01 [$-$0.01, 0.03]). The best candidate for the 7B agent comes from Qwen3.8-flash's proposals in 34\% of trajectories, the privileged action in 29\%, the 7B model's proposals in 24\%, Llama's in 14\%, and the random action in 9\% (sources overlap); for Llama, from the 7B model's and Qwen3.8-flash's proposals in 46\% each, the privileged action in 20\%, and Llama's own in 7\%. On the Llama pool, too, an executor's own proposals are not better for it (interaction $-$0.025 [$-$0.048, $-$0.003]). Both privileged and flash screens lose more than greedy replay on the Llama pool (0.26 and 0.17 at five rollouts, against 0.13). The 14B executor ran on a second server whose environment matched the first in every registered check (instruction list, products and prices, search fingerprint, data hashes; only the spaCy patch version differed, 3.8.18 versus 3.8.16); on the 7B pool's branches it reaches full reward in 20\% of rollouts (mean reward 0.52), and candidates from every source have non-negative mean net gain for it.

\begin{table*}[!htbp]
  \caption{WebShop, unscreened pools (60 trajectories each): differences in net reward gain (primary) and net success gain (secondary), nonparametric 90\% intervals. ``Other model'': deployed agent's own selection minus the other deployed model's, five rollouts per candidate each. ALFWorld-style: only LLM-proposed, privileged, and random candidates, as in the ALFWorld pools. LLM-proposed: candidates proposed by at least one LLM. $^\dagger$The success-based positive control of the Llama pool falls just short of 0.10, so its success readouts are descriptive only.}
  \label{tab:webshop}
  \centering
  \footnotesize
  \setlength{\tabcolsep}{3.4pt}
  \begin{tabular}{@{}lcc|cc@{}}
    \toprule
    & \multicolumn{2}{c|}{Deployed Qwen2.5-7B} & \multicolumn{2}{c}{Deployed Llama-3.1-8B} \\
    Contrast & Reward & Success & Reward & Success$^\dagger$ \\
    \midrule
    Positive control: matched $-$ random & 0.163 [0.128, 0.200] & 0.169 [0.114, 0.229] & 0.314 [0.258, 0.373] & 0.149 [0.090, 0.210] \\
    Matched $-$ greedy replay (P1w; H3w) & 0.061 [0.036, 0.086] & 0.040 [0.014, 0.069] & 0.131 [0.083, 0.181] & 0.094 [0.055, 0.136] \\
    Other model, same budget (H1w; H2w) & 0.081 [0.046, 0.114] & 0.075 [0.042, 0.112] & 0.166 [0.117, 0.217] & 0.080 [0.040, 0.124] \\
    Matched $-$ privileged action, earliest (S1w) & 0.144 [0.100, 0.190] & 0.161 [0.094, 0.230] & 0.248 [0.190, 0.305] & 0.094 [0.040, 0.156] \\
    \midrule
    ALFWorld-style: matched $-$ greedy & 0.057 [0.032, 0.083] & 0.049 [0.021, 0.082] & 0.088 [0.057, 0.124] & 0.073 [0.040, 0.110] \\
    ALFWorld-style: other model & 0.085 [0.051, 0.116] & 0.088 [0.048, 0.135] & 0.147 [0.103, 0.193] & 0.077 [0.039, 0.116] \\
    LLM-proposed: matched $-$ greedy & 0.051 [0.025, 0.078] & 0.034 [0.005, 0.070] & 0.070 [0.040, 0.102] & 0.052 [0.026, 0.081] \\
    LLM-proposed: other model & 0.045 [0.017, 0.073] & 0.050 [0.015, 0.094] & 0.125 [0.081, 0.171] & 0.053 [0.019, 0.090] \\
    \midrule
    14B, same budget (P2w) & 0.073 [0.046, 0.101] & 0.030 [0.003, 0.058] & --- & --- \\
    ALFWorld-style: 14B & 0.069 [0.042, 0.098] & 0.028 [0.003, 0.056] & --- & --- \\
    LLM-proposed: 14B & 0.053 [0.030, 0.078] & 0.013 [$-$0.006, 0.033] & --- & --- \\
    \bottomrule
  \end{tabular}
\end{table*}

\subsection{\texorpdfstring{$\tau^2$}{tau2}-bench}
\label{app:tau2}

\begin{table}[!htbp]
  \caption{\tautwo{}, Qwen3.5-2B deployed (115 failed conversations, 96 tasks). Net gain of the selected repair for the 2B agent; matched selection uses five of its ten rollouts per branch and is scored on the other five. 90\% intervals from a bootstrap over tasks, stratified by domain. Sensitivity: an admissibility rule without the authentication condition; and dropping the six conversations with unscored rollouts.}
  \label{tab:tau2}
  \centering
  \footnotesize
  \setlength{\tabcolsep}{3pt}
  \resizebox{\columnwidth}{!}{%
  \begin{tabular}{@{}lccc@{}}
    \toprule
    & Main & Other rule & Without 6 \\
    \midrule
    Matched $-$ random & 0.099 [0.067, 0.133] & 0.107 & 0.102 \\
    Matched $-$ greedy (P1$\tau$) & 0.016 [0.000, 0.033] & 0.017 & 0.020 \\
    Own $-$ 4B selection (P2$\tau$) & 0.026 [0.005, 0.048] & 0.035 & 0.028 \\
    \bottomrule
  \end{tabular}}
\end{table}

\Cref{tab:tau2} gives the pre-specified readouts. The positive control requires the lower bound of matched minus random selection to exceed 0.10; it does not, in the main analysis or in either sensitivity analysis, so we report P1$\tau$ and P2$\tau$ as bounds and not as confirmations. Both are nominally positive ($p=0.049$ and $0.021$), and both upper bounds are below 0.05. By domain, P2$\tau$ is 0.034 in airline and 0.021 in retail. With three rollouts per candidate on both sides, the loss is 0.025 when the 4B model selects for the 2B agent and 0.021 in the reverse direction.

The value of the selected repair grows little with budget: 0.160 with one 2B rollout per candidate, 0.189 with five, 0.173 with greedy replay, 0.164 with five 4B rollouts, 0.090 for random selection, and 0.280 for the best candidate in hindsight. Candidates that are tool calls carry almost no selection value (positive control 0.010); the effects come from text turns. Of the 7,330 sampled 2B rollouts, 21 exceeded the context limit, as did five greedy and two 4B rollouts; the main analysis counts them as failures, a rule fixed before P1$\tau$ and P2$\tau$ were computed. A pre-specified test of whether the loss at a position grows with the divergence between the two models' next-action distributions was inconclusive, because position-level losses were too unreliable to rank (split-half reliability below 0.20).

\paragraph{Loss shares.}
The shares in \cref{tab:ratio} are ratios of means, with intervals from 10{,}000 paired resamples of trajectories; for the 7B agent with Llama as selector, whose per-trajectory losses were not stored with the pool's, the interval scales the loss interval by the point estimate of the selection value. A simplified re-implementation with fixed halves gives shares within 0.14 of those in the table. Within pools, trajectories with more selection value tend to lose more to a substitute: the rank correlation is positive in six of seven pool--selector pairs (0.13--0.37, five with intervals excluding zero) and negative in one ($-0.24$, the 7B model selecting for Llama in ALFWorld).

\subsection{Post-Repair Success Objective}
\label{app:post}

Under the secondary objective, which credits a repair with its post-repair success probability $p_{e_0}(h_t,a')$ rather than its net gain, the conclusions about small budgets, greedy replay, the solvability oracle, LLM judges in ALFWorld, and other executors are unchanged; environment heuristics in ALFWorld fare better, as discussed below. On the unscreened ALFWorld pools, greedy replay loses 0.10 [0.07, 0.14] (7B) and 0.07 [0.04, 0.10] (Llama), and substitute executors lose 0.29 [0.23, 0.35] (14B for 7B), 0.06 [0.03, 0.09] (Llama for 7B), and 0.08 [0.05, 0.12] (7B for Llama). On the screened pools, relative to matched selection (post-repair success 0.72 in the gridworld), greedy replay loses 0.19 [0.12, 0.25], three rollouts lose 0.10 [0.06, 0.14], temperature 0.3 loses 0.02 [$-$0.03, 0.08], and the solvability oracle loses 0.47 [0.40, 0.54], more than under net gain because under this objective it ties across solvable repairs at every position. In ALFWorld (matched 0.97), greedy replay loses 0.14 [0.08, 0.20] and three rollouts 0.05 [0.02, 0.07]. With ten rollouts per candidate on each side, selecting by the 14B model's or Qwen3.8-flash's post-repair success instead of the deployed agent's loses 0.29 [0.21, 0.37] and 0.62 [0.55, 0.69] in ALFWorld, and 0.24 [0.17, 0.31] and 0.43 [0.37, 0.49] in the gridworld (trajectory bootstrap; the corresponding net-gain losses are 0.29, 0.63, 0.06, and 0.16). \Cref{tab:post} lists all evaluators on the screened pools under this objective.

\begin{table}[!htbp]
  \caption{Selection loss on the screened pools under the post-repair success objective (bias-corrected 90\% intervals). Matched selection reaches post-repair success 0.72 (gridworld) and 0.97 (ALFWorld). The acceptance-based published rules are omitted because they have no natural post-repair value when they decline to repair.}
  \label{tab:post}
  \centering
  \resizebox{\columnwidth}{!}{%
  \begin{tabular}{@{}lcc@{}}
    \toprule
    Evaluator & Gridworld & ALFWorld \\
    \midrule
    Random (positive control) & 0.51 [0.43, 0.58] & 0.74 [0.68, 0.78] \\
    Deployed agent, 3 rollouts & 0.10 [0.06, 0.14] & 0.05 [0.02, 0.07] \\
    Deployed agent, 5 rollouts & 0.08 [0.04, 0.12] & 0.02 [$-$0.00, 0.04] \\
    Deployed agent, 10 rollouts & 0.02 [$-$0.02, 0.05] & 0.01 [$-$0.02, 0.02] \\
    Same model, temperature 0.3 & 0.02 [$-$0.03, 0.08] & --- \\
    Greedy replay & 0.19 [0.12, 0.25] & 0.14 [0.08, 0.20] \\
    Solvability oracle $F$ & 0.47 [0.40, 0.54] & 0.74 [0.68, 0.78] \\
    Oracle, ties broken by slack & 0.37 [0.27, 0.46] & 0.07 [0.01, 0.14] \\
    Planner action, earliest position & --- & 0.06 [0.01, 0.13] \\
    Reasoning LLM judge & 0.02 [$-$0.03, 0.08] & 0.48 [0.36, 0.61] \\
    Informed LLM judge & --- & 0.11 [0.04, 0.20] \\
    Deployed model judging itself & 0.49 [0.39, 0.58] & 0.72 [0.62, 0.82] \\
    \bottomrule
  \end{tabular}}
\end{table}

Two differences from the net-gain results deserve mention. In ALFWorld, environment heuristics come much closer to matched selection under this objective: the planner's action at the earliest position loses 0.06 and the oracle with slack tie-breaking 0.07, against 0.14 and 0.28 under net gain. The earliest position is also where the original action most often succeeds anyway (0.19 on average, against 0.08 at the latest position), so a repair placed there scores well on post-repair success while adding little; what the heuristics miss is mainly where a repair \emph{adds} success. In the gridworld, the oracle with slack tie-breaking loses 0.37, because under this objective it ranks candidates by absolute rather than gained slack, which favors early positions; the reasoning judge, by contrast, stays within the margin.

\subsection{Unscreened Gridworld Failures}
\label{app:gridpool}

On the first 100 failures on new gridworld maps, including timeouts before the door, we labeled all alternatives at four positions with a two-stage design: five rollouts per branch select three candidates per trajectory, which receive 24 more. Selecting with five deployed-agent rollouts beats greedy replay by 0.05 [0.02, 0.08] and the solvability oracle by 0.07 [0.04, 0.10], 20--40\% of the screened-pool losses, and is level with the oracle with slack tie-breaking. The effects are smaller because matched selection uses five rollouts rather than twenty and many unscreened gridworld failures have no useful single-step repair: only 62\% of trajectories have a candidate with estimated net gain above 0.10. The availability of useful single-step repairs, and where the best one lies, thus depend on the environment and the executor; this does not show whether a failure has a single decisive error.

\section{Screened Pools and Readers That Do Not Execute}
\label{app:readers}

\subsection{Screened Pools}
\label{app:screened}

This section gives the full account behind the summary in \cref{sec:main}.

\begin{table}[!htbp]
  \caption{Selection loss on the screened pools, relative to matched selection with all of the deployed agent's rollouts (20 per fold in the gridworld, 15 in ALFWorld; deployment net gain 0.662 and 0.851). Larger is worse; bias-corrected 90\% intervals. The first block keeps the target and changes the budget or the decision rule; the published rules use the deployed agent's own rollouts, so their losses come from the rule, not from mismatch. The second block changes the target. Other executors are compared at equal budget in \cref{fig:forest}.}
  \label{tab:screened}
  \centering
  \footnotesize
  \setlength{\tabcolsep}{2pt}
  \resizebox{\columnwidth}{!}{%
  \begin{tabular}{@{}lcc@{}}
    \toprule
    Evaluator & Gridworld (34) & ALFWorld (30) \\
    \midrule
    Random (pos.\ control) & 0.66 [0.57, 0.74] & 0.75 [0.69, 0.81] \\
    \midrule
    \multicolumn{3}{@{}l}{\emph{Same target, smaller budget or another rule}} \\
    Deployed, 3 rollouts & 0.11 [0.06, 0.16] & 0.07 [0.03, 0.11] \\
    Deployed, 10 rollouts & 0.01 [$-$0.02, 0.04] & 0.02 [$-$0.02, 0.05] \\
    2-of-3 acceptance & 0.32 [0.25, 0.39] & 0.23 [0.16, 0.30] \\
    Single-replay flip & 0.43 [0.36, 0.50] & 0.28 [0.22, 0.35] \\
    Earliest single success & 0.58 [0.48, 0.68] & 0.25 [0.18, 0.34] \\
    Latest significant step & $-$0.01 [$-$0.04, 0.03] & 0.19 [0.11, 0.28] \\
    \midrule
    \multicolumn{3}{@{}l}{\emph{Different target}} \\
    Same model, temp.\ 0.3 & 0.03 [$-$0.03, 0.09] & --- \\
    Greedy replay & 0.23 [0.15, 0.30] & 0.18 [0.10, 0.25] \\
    Solvability oracle $F$ & 0.19 [0.12, 0.26] & 0.75 [0.69, 0.81] \\
    Oracle, slack tie-break & 0.04 [$-$0.02, 0.09] & 0.28 [0.20, 0.36] \\
    Planner action, earliest & --- & 0.14 [0.06, 0.23] \\
    Reasoning judge & 0.06 [$-$0.00, 0.13] & 0.53 [0.41, 0.66] \\
    \quad without reasoning & 0.25 [0.14, 0.38] & 0.42 [0.30, 0.55] \\
    \quad informed (contents) & --- & 0.19 [0.10, 0.28] \\
    Deployed model as judge & 0.59 [0.48, 0.70] & 0.74 [0.64, 0.83] \\
    \bottomrule
  \end{tabular}}
\end{table}

\paragraph{Greedy replay.}
Deterministic replay loses 0.23 in the gridworld and 0.18 in ALFWorld. Part of the reason is that greedy decoding repeats its own mistakes (after bumping into a wall or a closed door it repeats the action 39\% of the time, against 33\% under sampling); in a pre-specified follow-up, switching to sampling whenever the greedy agent revisits a state recovers 0.10 of the 0.23 gridworld loss. The remainder reflects which branches the deterministic policy happens to solve, which differs from the stochastic policy that will actually run.

\paragraph{Environment signals know what, not where.}
Solvability loses 0.19 in the gridworld: at the irreversible point several actions usually restore solvability, and choosing among them at random rather than with the deployed agent's rollouts loses 0.31 [0.22, 0.40]. In ALFWorld every candidate branch leaves the task solvable, so solvability equals random selection. Breaking ties by slack nearly closes the gap in the gridworld (0.04) but not in ALFWorld (0.28), and the planner's action at the earliest position still loses 0.14 (0.06 under the post-repair objective; \cref{app:post}). The planner usually knows \emph{what} to change---in 29 of 30 screened ALFWorld trajectories the best candidate is the planner's action---but not \emph{where} the deployed agent benefits most, which depends on how the agent behaves after the intervention; across the three positions of a trajectory, the best net gain differs by 0.51 on average.

\paragraph{Published decision rules.}
Applied to the deployed agent's own rollouts, acceptance rules---two of three replays succeed \citep{ma2025dover}, or a single replay flips the outcome \citep{bonagiri2026causalflow}---do not rank the interventions they accept, and many accepted repairs succeeded by chance (losses 0.23--0.43). The earliest position at which a single replay succeeds \citep{zhang2025agentracer} loses 0.58 in the gridworld, where early single successes are often luck. The latest step with a significant effect under same-policy resampling \citep{shah2026car} is nearly lossless in the gridworld, where traps make the last solvable step decisive, but loses 0.19 in ALFWorld, which has no irreversible actions.

\paragraph{Readers that do not execute.}
In the gridworld, a reasoning judge that reads the log loses only 0.06, indistinguishable from the oracle with slack tie-breaking. Removing the executor description leaves 26 of 34 choices unchanged, whereas disabling reasoning costs 0.19 [0.10, 0.31]; this is consistent with the judge relying on information about the environment rather than about the executor. In ALFWorld, where the log does not reveal the contents of closed containers, the judge loses 0.53. Given the initial contents of every container, it improves by 0.35 [0.23, 0.46] and picks the planner's action in all 30 trajectories, but still loses 0.19 [0.10, 0.28] because it chooses the best position in only 47\% of them (see also \cref{app:extra}). In an exploratory study on the unscreened 7B pool, one human annotator who read each failed log and chose among the same candidates gained 0.34 [0.25, 0.42]: more than random selection (by 0.22), but less than half of matched selection (0.82) and below greedy replay (0.69) (\cref{app:human}).

\subsection{A Human Reader (Exploratory)}
\label{app:human}

To add a human to the readers that do not execute, one annotator---a co-author new to ALFWorld who had not seen any rollout result---chose a repair for each of the 60 unscreened 7B failures. The annotator saw the task, the initial description, the full failed log, and the three fixed positions with their candidates (shuffled within each position, sources hidden), and was told that the 7B model would continue from the replaced action at temperature 0.7: the same information the LLM proposers received. The written protocol explained the interface and the basic rules of the environment but gave no hints or examples. For each trajectory the annotator recorded a choice, a reason, and a confidence, and could mark the information as insufficient or name a decisive step; active annotation time was 4.4 hours in total. The readouts were drafted before annotation, but the protocol was not frozen after the pilot as planned, there is one annotator, and agreement cannot be assessed, so we report the study as exploratory and do not generalize to human readers.

\begin{table}[!htbp]
  \caption{A human reader on the 60 unscreened 7B ALFWorld failures: net gain for the 7B agent, scored on all ten of its rollouts per branch (paired trajectory bootstrap, 10,000 replicates, 90\% intervals). Rollouts: deployed-agent rollouts per trajectory used for selection.}
  \label{tab:human}
  \centering
  \footnotesize
  \setlength{\tabcolsep}{3pt}
  \resizebox{\columnwidth}{!}{%
  \begin{tabular}{@{}lcc@{}}
    \toprule
    Selection & Net gain & Rollouts \\
    \midrule
    Human reads the log (one annotator) & 0.335 [0.247, 0.423] & 0 \\
    Random candidate & 0.115 [0.099, 0.130] & 0 \\
    LLM screen (first flash proposal), $k=5$ & 0.270 [0.187, 0.357] & 30 \\
    Planner action, earliest position (privileged) & 0.682 [0.605, 0.757] & 0 \\
    Greedy replay & 0.687 [0.618, 0.754] & 21 \\
    Matched selection (5 rollouts) & 0.822 [0.762, 0.878] & 107 \\
    \midrule
    Matched $-$ human & 0.487 [0.395, 0.580] & \\
    Human $-$ greedy replay & $-$0.352 [$-$0.454, $-$0.251] & \\
    Human $-$ random & 0.220 [0.136, 0.306] & \\
    Human $-$ LLM screen ($k=5$) & 0.065 [$-$0.045, 0.179] & \\
    \bottomrule
  \end{tabular}}
\end{table}

\Cref{tab:human} gives the results. The annotator's repairs gain 0.34, more than random selection but less than half of matched selection and below greedy replay, which executes the branches; they are comparable to the LLM screen with five verification rollouts per survivor. The annotator chose the planner's action (without knowing it was one) in 27 trajectories, with a mean net gain of 0.66, against 0.07 in the other 33; this grouping is not randomized. The annotator marked the information as insufficient in 44 of 60 trajectories and left the optional decisive-step field blank in 54, so the study cannot speak to whether these failures have a single decisive error. One submission, which chose \texttt{inventory} with a note that it was testing the interface, may be accidental; excluding it changes the human's net gain to 0.339 and matched minus human to 0.483. Scored on other executors, the same choices gain 0.28 for the 14B model and 0.36 for Llama (descriptive; the task asked for repairs for the 7B agent).

\subsection{Reading versus Running}
\label{app:extra}

The same model can act as a judge or as an executor. With reasoning enabled, Qwen3.8-flash chooses better repairs for the 7B agent by reading the log than by running the branches itself (gridworld 0.11 [0.04, 0.18]; ALFWorld 0.11 [$-$0.03, 0.26]). This is confounded by reasoning, which the judge uses and the executor does not: without it, the judge loses 0.25 in the gridworld, more than Qwen3.8-flash's rollouts (0.17), though less in ALFWorld (0.42 versus 0.64). The deployed 7B model chooses far better by running than by reading (by 0.59 and 0.72).

\section{Training Version and Per-Position Labels}
\label{app:version}

\subsection{Training Version}
\label{app:training}

\paragraph{Successors.}
Training positions come from training episodes of the 7B agent that share no game with the pool. At each of 1,116 decision points, candidates (the planner's action, two proposals each from the 7B agent and Qwen3.8-flash, and a random action) received three 7B rollouts; the label is the best candidate at the 558 points with the highest estimated gain, written with a template thought. Each label is paired with one of the agent's own steps from a successful episode. Training uses LoRA with rank 16, $\alpha=32$, two epochs, and learning rate $10^{-4}$; the two successors of a pair differ only in the seed. Successor rollouts on the pool use the same environment seeds for all successors.

For the first pair, 89\% of the labels are the planner's action. The second pair changes one thing: the label at each point is the best candidate among those a pipeline without a planner would have, namely the candidates proposed by a model and the random one. The same 1,116 points, label rollouts, selection rule, paired steps, and training recipe are used, and before training we required the selection and export code to reproduce the first pair's training file exactly. The resulting labels share 66\% of their decision points with the first pair's; 44\% are still the planner's action, because a model also proposed it, and their estimated gain is lower (0.61 against 0.95 in the label rollouts). The plan for the second pair, written before its labels were generated, named one test (N1: the loss when the successor selects for the original agent has a 90\% lower bound above 0.10), conditional on the successor having changed (N2: its own success from the pool's positions rises, lower bound above 0). It predicted that both would be confirmed with a loss between 0.35 and 0.60, and that the positive control in the reverse direction would fail for both seeds; the last prediction was wrong for one seed.

\begin{table*}[!htbp]
  \caption{Selection across one round of training, unscreened 7B ALFWorld pool (60 trajectories; five rollouts per candidate on both sides; 50 random splits; 90\% intervals over trajectories). Shares divide the loss by matched minus random selection. $\Delta$: preference for the deployed agent's choice over the selector's, on held-out rollouts, where the two choices differ.}
  \label{tab:training}
  \centering
  \footnotesize
  \setlength{\tabcolsep}{2.5pt}
  \resizebox{\textwidth}{!}{%
  \begin{tabular}{@{}lcccc@{}}
    \toprule
    & \multicolumn{2}{c}{First pair (labels include the planner's candidates)} & \multicolumn{2}{c}{Second pair (labels without them)} \\
    & Seed 1 & Seed 2 & Seed 1 & Seed 2 \\
    \midrule
    Own success from the pool's positions (original agent: 0.116) & 0.456 & 0.359 & 0.308 & 0.210 \\
    \midrule
    \multicolumn{5}{@{}l}{\emph{Deployed: original 7B agent; selector: successor} (matched $-$ random: 0.711 [0.658, 0.759])} \\
    Loss & 0.598 [0.527, 0.668] & 0.586 [0.521, 0.653] & 0.557 [0.483, 0.634] & 0.548 [0.476, 0.623] \\
    Share & 0.84 & 0.82 & 0.78 & 0.77 \\
    Loss, selector unsaturated ($n$) & 0.503 (25) & 0.550 (33) & 0.476 (34) & 0.517 (44) \\
    $\Delta$ for the 7B agent & 0.662 [0.588, 0.735] & 0.639 [0.570, 0.705] & 0.625 [0.551, 0.699] & 0.608 [0.535, 0.681] \\
    $\Delta$ for the successor & $-$0.037 [$-$0.088, 0.013] & $-$0.025 [$-$0.067, 0.019] & $-$0.059 [$-$0.105, $-$0.012] & $-$0.005 [$-$0.051, 0.041] \\
    \midrule
    \multicolumn{5}{@{}l}{\emph{Deployed: successor; selector: original 7B agent}} \\
    Matched selection, net gain & 0.121 [0.069, 0.171] & 0.102 [0.050, 0.154] & 0.181 [0.119, 0.243] & 0.151 [0.099, 0.201] \\
    Matched $-$ random & 0.128 [0.088, 0.166] & 0.104 [0.065, 0.143] & 0.165 [0.115, 0.211] & 0.125 [0.085, 0.166] \\
    Loss & 0.043 [0.004, 0.082] & 0.030 [$-$0.004, 0.064] & 0.054 [0.022, 0.088] & 0.019 [$-$0.015, 0.054] \\
    Loss, other seed selects & 0.045 [0.006, 0.085] & 0.023 [$-$0.002, 0.049] & 0.041 [0.005, 0.080] & 0.020 [$-$0.009, 0.051] \\
    Loss, 7B restricted to matched position & $-$0.003 [$-$0.029, 0.023] & $-$0.008 [$-$0.032, 0.019] & 0.007 [$-$0.014, 0.030] & $-$0.010 [$-$0.030, 0.014] \\
    Loss, 7B position, own action & 0.060 [0.026, 0.094] & 0.046 [0.018, 0.076] & 0.070 [0.038, 0.103] & 0.048 [0.016, 0.081] \\
    \midrule
    \multicolumn{5}{@{}l}{\emph{Position by position (142 positions worth repairing for the original agent)}} \\
    Not worth repairing for the successor & 0.82 [0.77, 0.87] & 0.78 [0.71, 0.85] & 0.73 [0.67, 0.80] & 0.75 [0.69, 0.82] \\
    Successor's best candidate $-$ old label & $-$0.011 [$-$0.034, 0.015] & $-$0.018 [$-$0.042, 0.010] & 0.004 [$-$0.019, 0.026] & $-$0.014 [$-$0.033, 0.004] \\
    \bottomrule
  \end{tabular}}
\end{table*}

\paragraph{Selection loss.}
\Cref{tab:training} gives both directions. The implementation follows \cref{sec:exp} with one difference: a split assigns the same rollout indices to selection and evaluation for every branch, and a selector uses only the selection indices, so that the environment seeds the successors share never link a selector's choice to the deployed model's evaluation. On the original agent it reproduces the pre-specified analysis (matched selection 0.826 against 0.82; greedy loss 0.139 against 0.13), and recomputing the first pair's four losses with the pre-specified code, which splits each branch independently, changes none by more than 0.006. For one successor of the second pair the reverse direction clears the positive control (lower bound 0.115), and the original agent's choices cost it 0.054, a third of what selection is worth to it. ``Unsaturated'' follows the stratification of \cref{sec:why}: the selector, keeping the original action at the position of the deployed agent's best repair, succeeds less than half the time. When the successor is deployed, restricting the 7B agent's choice to the position that matched selection picked removes the loss, and letting the 7B agent pick only the position leaves it; the two interventions are not an additive decomposition, but both point to the position as what the old choice gets wrong.

\paragraph{Labels position by position.}
Call a position worth repairing for a policy when its cross-fitted best candidate beats what the policy does on its own from that position by at least 0.20 (five rollouts select, five evaluate, both directions averaged). The original agent is worth repairing at 142 of the 180 positions. Of these, 73--82\% are not worth repairing for a successor (\cref{tab:training}). As a same-policy baseline, of the positions worth repairing in one half of the original agent's rollouts, 0.04 are not in its other half; replacing the second half by a successor's rollouts gives 0.44--0.51. At the positions worth repairing for the original agent, the successor's own cross-fitted best candidate does not beat the original agent's label in the successor's success probability (\cref{tab:training}). Restricting the candidates to those a pipeline without a planner would have leaves the picture unchanged for all four successors (0.77--0.84 not worth repairing at 92 positions; best candidate minus old label between $-0.04$ and 0.00). These per-position readouts were pre-specified in separate dated plans; the values here use an inclusive threshold throughout.

\paragraph{The same estimator when nothing changed.}
On \tautwo{} we trained successors of Qwen3.5-4B in the same way, with two-fold cross-fitting over tasks, 800 positions of the pool described in \cref{app:setup}, and six rollouts per branch. The successors did not measurably change: continuing on their own they succeed 0.638 of the time against 0.641 before ($-0.004$ [$-0.017$, 0.010]). The cross-fitted estimator nevertheless reports 0.52 [0.45, 0.59] of the original labels as no longer worth repairing. Two halves of the unchanged policy's rollouts disagree on 0.50 of the positions worth repairing, and replacing the second half by the successor's gives 0.48. With three rollouts per half and gains near the threshold, half of the labels do not replicate, whoever the second policy is.

\subsection{Per-Position Labels Across Executors}
\label{app:labels}

\begin{table}[!htbp]
  \caption{Per-position labels (post hoc). A selector labels a position with its best candidate when that candidate's net gain in its own rollouts is at least 0.20. Labels: the selector's number of labels / the deployed agent's own. Same: share of the selector's labels that coincide in position and action with the deployed agent's. Gain: mean net gain of the selector's labels for the deployed agent, against the held-out gain of the deployed agent's own labels (Own); each mean is over the positions that selector labels. WebShop gains are rewards.}
  \label{tab:labels}
  \centering
  \footnotesize
  \setlength{\tabcolsep}{3pt}
  \resizebox{\columnwidth}{!}{%
  \begin{tabular}{@{}llcccc@{}}
    \toprule
    Pool & Deployed $\leftarrow$ selector & Labels & Same & Gain & Own \\
    \midrule
    ALFWorld & 7B $\leftarrow$ 14B & 127 / 150 & 0.86 & 0.677 & 0.717 \\
    ALFWorld & 7B $\leftarrow$ Llama & 148 / 150 & 0.90 & 0.734 & 0.717 \\
    ALFWorld & Llama $\leftarrow$ 7B & 154 / 168 & 0.92 & 0.776 & 0.803 \\
    WebShop & 7B $\leftarrow$ 14B & 76 / 66 & 0.34 & 0.100 & 0.165 \\
    WebShop & 7B $\leftarrow$ Llama & 63 / 66 & 0.29 & 0.099 & 0.165 \\
    WebShop & Llama $\leftarrow$ 7B & 73 / 80 & 0.48 & 0.227 & 0.321 \\
    \tautwo{} & 2B $\leftarrow$ 4B & 49 / 76 & 0.61 & 0.352 & 0.266 \\
    \midrule
    ALFWorld & 7B $\leftarrow$ successors, first pair & 66, 59 / 150 & 0.47, 0.49 & 0.286, 0.295 & 0.717 \\
    ALFWorld & 7B $\leftarrow$ successors, second pair & 72, 71 / 150 & 0.43, 0.41 & 0.319, 0.347 & 0.717 \\
    \bottomrule
  \end{tabular}}
\end{table}

\Cref{tab:labels} labels every position separately. In ALFWorld the planner's action is the best candidate at most positions for every executor, and labels chosen by another model are worth nearly as much as the agent's own; the selection loss of \cref{sec:main} arises when one repair must be chosen among a trajectory's positions. In WebShop the labels agree in under half of the cases and are worth about a third less. On \tautwo{} the 4B model labels fewer positions, and those labels are worth no less to the 2B agent than its own. The successors' labels fall on 53--68 of the 150 positions that the 7B agent needs repaired; 42--49\% of their labels gain nothing for it, and the 7B agent's own labels are worth 0.06--0.10 to the successors.

\section{Low-Cost Verification}
\label{app:method}

\begin{figure}[!htbp]
  \centering
  \includegraphics[width=\columnwidth]{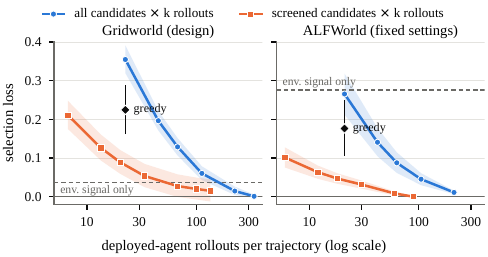}
  \caption{Screened pools: selection loss against the number of deployed-agent rollouts per trajectory, when every candidate receives $k$ rollouts (blue) or only the candidate preferred by the environment signal at each position does (orange). Shaded: 90\% bootstrap intervals over trajectories. Settings were fixed on the gridworld and applied unchanged to ALFWorld.}
  \label{fig:budget}
\end{figure}

\Cref{fig:budget} compares the privileged screen of \cref{sec:method} with spending the same budget on all candidates. In screened ALFWorld, ten rollouts per survivor---60 rollouts per trajectory, 19\% of what matched selection uses---come within 0.009 [0.004, 0.014] of matched selection. At a budget comparable to greedy replay (18 versus 21 rollouts), the screen beats greedy replay by 0.13 [0.06, 0.20] and one rollout per candidate by 0.22 [0.17, 0.26]. It also beats the environment signal alone by 0.27 [0.20, 0.34], because the signal cannot choose the position. In the gridworld, where the environment signal alone is already close to matched selection, the procedure matches it (0.01 [$-$0.02, 0.05]); on the unscreened gridworld pool it improves on five-rollout matched selection by 0.02 at 29\% of the budget, by removing noisy candidates from the maximization. On the unscreened ALFWorld pool, the flash screen fails because its survivors contain the deployed agent's best candidate in 22\% of trajectories, against 81\% for the planner screen.

\begin{table}[!htbp]
  \caption{The LLM-located, verified pipeline of \cref{sec:method} and comparisons on the same 60 unscreened 7B ALFWorld failures (deployment net gain; nonparametric 90\% intervals). Rollouts: deployed-agent rollouts per trajectory used for selection. The last three rows select among the fixed candidates of \cref{sec:exp} at three positions.}
  \label{tab:pipeline}
  \centering
  \footnotesize
  \setlength{\tabcolsep}{3pt}
  \resizebox{\columnwidth}{!}{%
  \begin{tabular}{@{}lcc@{}}
    \toprule
    Selection & Net gain & Rollouts \\
    \midrule
    LLM locates and proposes; 2-of-3 check & 0.32 [0.22, 0.40] & 6 \\
    \quad no check (first proposal) & 0.22 [0.14, 0.30] & 0 \\
    \quad greedy check (first greedy success) & 0.31 [0.22, 0.40] & 2 (greedy) \\
    \quad more successes than the original action & 0.32 [0.23, 0.40] & 9 \\
    \quad better proposal in hindsight & 0.37 [0.29, 0.46] & --- \\
    LLM screen (first proposal per position), $k=1$ / 5 & 0.24 / 0.27 & 6 / 30 \\
    Greedy replay, LLM-proposed / all candidates & 0.52 / 0.69 & 17 / 21 (greedy) \\
    Matched selection, LLM-proposed / all candidates & 0.64 / 0.82 & 83 / 107 \\
    \bottomrule
  \end{tabular}}
\end{table}

\paragraph{The LLM-located pipeline.}
In a follow-up specified before running it (exploratory), an LLM both locates and proposes: Qwen3.8-flash with reasoning read each of the 60 unscreened failed 7B logs, named one step, and proposed two alternatives there; the deployed agent ran each three times, and the more successful one was applied if it succeeded at least twice \citep{ma2025dover}. Scored on ten fresh rollouts per branch, the pipeline gains 0.32 [0.22, 0.40] with six rollouts per trajectory, against 0.22 [0.14, 0.30] for applying the first proposal unverified. Matched selection over the LLM-proposed candidates at three fixed positions, a different candidate set, gains 0.64 with 83 rollouts. The pipeline's gap lies mostly in its candidates: the better of its two proposals in hindsight is worth only 0.37, and the pipeline does well mainly when the LLM infers from the log where the target object is (0.63 when a proposal is the planner's action, 0.16 otherwise; post hoc). Its verification task also differs from ranking a fixed candidate pool: with two candidates at one position, one greedy replay verifies as well as three samples (0.31), because verification only has to separate clearly good from clearly bad proposals, whereas greedy replay loses 0.13 when it ranks the full candidate set (\cref{tab:pipeline}).
The LLM's chosen step is early (median step 5 of 50; 48 of 60 in the first third) and coincides with one of the three fixed positions in only 3 trajectories. All 60 locations and all 120 proposals were valid; a proposal was the planner's action in 20 of the 58 trajectories where that action differs from the original. Across proposals, the three verification rollouts succeeded 0 times for 65 proposals and 3 times for 28, and only 27 proposals fell in between, which is why a single greedy replay verifies as well as three samples here. At least one proposal was accepted in 27 trajectories, 14 of them with the planner's action, and accepted repairs gained 0.70 on average. The two calls to Qwen3.8-flash per trajectory used about 5,500 input and 2,400 output tokens in total; this cost is not included in the rollout counts.

\paragraph{Three rollouts in WebShop.}
Three rollouts per candidate reach 0.13 of the 0.15 in reward that five reach on the 7B pool and 0.26 of 0.31 on the Llama pool.

\paragraph{The privileged screen: details.}
The screen keeps, at each position, the candidate preferred by a privileged signal (solvability with slack tie-breaking, or the planner's or privileged action). The procedure was fixed on the gridworld and applied unchanged elsewhere; because it was motivated by an ALFWorld observation, the unscreened pools provide its test. In WebShop, on the 7B pool the screen (0.044 [0.015, 0.075] at five rollouts) is no better than screening with Qwen3.8-flash's first proposal (0.059), and on the Llama pool it loses 0.26 [0.20, 0.32], more than greedy replay (0.13).

\end{document}